\documentstyle[a4,12pt,cite,epsfig]{article}
\begin{document}
\renewcommand{\arraystretch}{1.6} 
\begin{titlepage}
%
%
\begin{flushright}
{\large\bf DESY 00-118}\\
\vspace{2mm}
\end{flushright}
\vspace{8.mm}

\begin{center}
{\Large\bf M{\o}ller Scattering Polarimetry\\ 
\vspace{0.7mm}

for\\ 

\vspace*{1.9mm}

High Energy ${\mathbf e^+ e^-}$ Linear Colliders}\\
\end{center}
\vspace{0.4cm}
 
\begin{center}
{\large G}{\small IDEON} {\large A}{\small LEXANDER}\footnote{On 
Sabbatical leave from Tel-Aviv University.}$^,$\footnote{
e-mail: alex@lep1.tau.ac.il.}\\ 
Institut f\"{u}r Physik\\
Humboldt-Universit\"{a}t zu Berlin, Germany\\
11015 Berlin, Germany\\ 
\vspace{2mm}

and \\
\vspace{2mm}

{\large I}{\small ULIANA} {\large C}{\small OHEN}\footnote{ 
e-mail: cohen@lep1.tau.ac.il}
 
School of Physics and Astronomy\\
Raymond and Beverly Sackler Faculty of Exact Sciences\\
Tel-Aviv University,
Tel-Aviv 69978, Israel\\
 \end{center}
\vspace{0.7cm}

\begin{abstract}

\noindent
The general features of the M{\o}ller scattering  
and its use as an electron polarimeter are described and studied
in view of the planned future high energy $e^+ e^-$ linear 
colliders. In particular the study concentrates on the TESLA collider  
which is envisaged to operate with longitudinal 
polarised beams at a centre of mass energy of
the order of 0.5 TeV with a luminosity of about 
${\cal L}$ = $10^{34}~{\mathrm cm^{-2}sec^{-1}}$.   

\end{abstract}
\end{titlepage}

\newpage
\section{Introduction}
\label{intro}
It is for some time that the high energy physics community
is of the opinion that in the near future there will be a need 
for the facility of a high energy linear $e^+ e^-$ collider with a nominal
energy around 0.5 TeV in the centre of mass (CM) system. A conceptual 
design of such a collider, known under the name TESLA, and its physics
program is described in some details in Ref. \cite{conceptual}.
It has further been pointed out that the option of 
longitudinal polarized electron beams in such high energy 
colliders, like TESLA, will enrich significantly 
the physics capabilities of the device \cite{casal}. 
The use of polarised beams requires however a continuous monitoring and
sufficient accurate measurement of the beam polarisation during the entire
collider operation.\\

\noindent
In addition to the widely used Compton scattering polarimeter, the
$e^-e^- \to e^-e^-$ M{\o}ller scattering process
has also been utilised to 
evaluate the polarisation level of the electron beams. 
Unlike the Compton polarimeter the operation of a M{\o}ller polarimeter may
need dedicated accelerator 
runs but its relatively simple construction and operation and
the large counting rates makes it nevertheless a rather attractive device.
Here one should note that the method applied for the M{\o}ller polarimeter
can also be applied almost without any change to the measurement 
of a positron beam polarisation
by replacing the M{\o}ller process with the Bhabha scattering \cite{bhabha}. \\
 
\noindent
Several colliders have in fact already used M{\o}ller polarimeters
to monitor their polarized electron beams.
The Stanford Linear Collider 
(SLC) has primarily used a precise Compton 
polarimeter to monitor the beam and measure the electron beam polarisation
\cite{swartz1,swartz2}. 
In addition it has also engaged two single-arm 
M{\o}ller polarimeter for beam polarisation diagnosis.
Many fixed target experiments, e.g. those described in 
references \cite{mol-e143,mol-e154,mol-jlab,mol-mit}, 
were running with polarised beams monitored by 
M{\o}ller polarimeters. 
Finally the M{\o}ller like scattering, $\mu e \to \mu e$, 
was used in the SMC experiment at CERN 
\cite{mol-muon} to measure the polarisation
of the muon beam.  
The M{\o}ller measurement can be also be carried out after the 
colliding beams interaction point (IP). 
From this point of view, the M{\o}ller polarimeters are more suited for 
the NLC, JLC or CLIC linear colliders, than for TESLA,  because of the 
non-zero crossing, the extraction 
does not require bending of the electron 
trajectories after the IP.\\  

\noindent
In this paper we describe the outcome of a detailed study 
which explored the feasibility and possibility to use the  
M{\o}ller scattering process as a method for the longitudinal
polarisation measurement of the TESLA electron beam. 
In Sections 2 and 3 we describe in some details the various properties 
of the M{\o}ller scattering, with and without polarised beams, and
also  review some of the technical characteristics of the  
M{\o}ller polarimeters used in recent high energy experiments
emphasizing the specific TESLA 
needs. 
The expected event rates and the effects of the energy deposition in 
the target
by the electron beam are dealt with in Section 4.
In Section 5 we consider two 
somewhat different methods for the beam polarisation 
measurement and evaluate their envisaged performance.  
\section{The M{\o}ller scattering}
\label{Isoscalar}
\subsection{The basic formulae}
\begin{figure}[hbtp]
\centerline{\epsfig{file=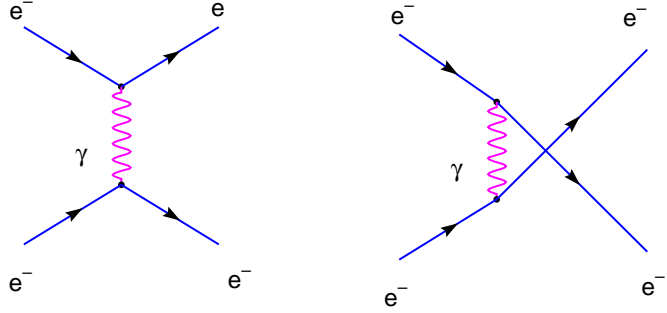,bbllx=10pt,
bblly=60pt,bburx=280pt,bbury=211pt,clip=}}
\caption{The lowest order Feynman diagrams describing the 
M{\o}ller elastic scattering.} 
\label{diagrams}
\end{figure}

\noindent
The lowest order $e^- e^- \to e^- e^-$   
M{\o}ller elastic scattering diagrams are the t-channel and u-channel
$\gamma$ exchanges\footnote{Despite the high energy of the beam in the
laboratory system,
the $\sqrt{s}$ is less than 1 GeV therefore the Z$^0$ gauge boson 
exchange contribution is negligible.} ~shown in Fig. \ref{diagrams}.
Each of these two diagrams contributes to 
the two possible spin configurations of the initial electrons, namely 
the parallel and anti-parallel states.
From the Fermi-Dirac statistics follows that the relative phase 
between the two diagrams is negative. This has important consequences
on the spin dependence of the cross section. For the anti-parallel spin 
configuration the scattered spins are also anti-parallel.
The anti-parallel spin state contains an additional negative phase
between the two possible orientations of the outgoing spins.
As a result the amplitudes
add and the cross section is larger for anti-parallel spin configuration
generating a non-zero asymmetry.\\

\noindent
The M{\o}ller differential elastic cross section at tree level and 
in the CM system is given by: 
\begin{equation}
\frac{d\sigma}{d\Omega}(s)\ = \ \frac{\alpha^2}{s}\frac{(3+\cos^2\theta)^2}
{\sin^4\theta} \left \{1 - P_L^BP_L^TA_L(\theta)-P_t^BP_t^TA_t(\theta)
\cos(2\phi-\phi_B-\phi_T) \right \} ~,
\label{mollerf}
\end{equation}   
for high $E_{CM}^2 = s$ values so that the electron mass squared
$m^2_e$ can be neglected.

\begin{figure}
\centering{\psfig{figure=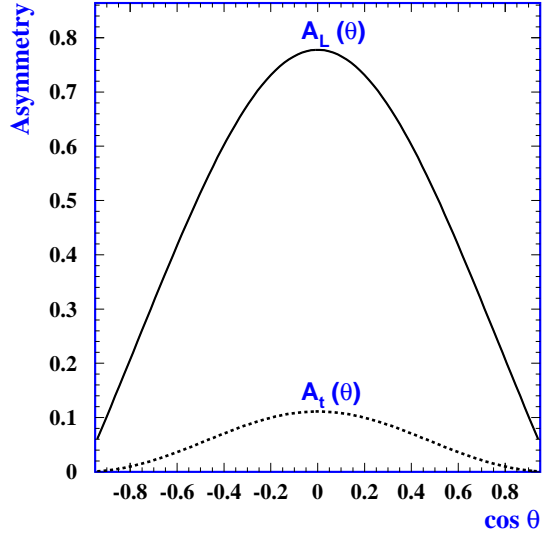,width=8cm,height=8cm}}
\caption{The longitudinal and the transverse asymmetry values
$A_L$ and $A_t$ are plotted as a function of
$\cos\theta$.}
\label{asyfun}
\end{figure}

\noindent
Here:\\
$\alpha$ = the fine structure constant at low energies 
which is equal to 1/137;\\
$\theta$ = the CM frame polar scattering angle;\\
$\phi$ = the CM azimuthal angle of the scattered electron;\\ 
$P_L^B$ and $P_L^T$ = longitudinal polarisation of the beam and target;\\
$P_t^B$ and $P_t^T$ = transverse polarisation of the beam and target;\\
$\phi_B$ and $\phi_T$ =  the azimuthal angles of the beam and target
transverse polarisation vectors.\\
\noindent
The longitudinal and transverse asymmetry functions,
$A_L(\theta)$ and $A_t(\theta)$, are defined as:\\
\begin{equation}
A_L(\theta)\ =\ \frac{(7+\cos^2\theta)\sin^2\theta}{(3+\cos^2\theta)^2}
\ \ \ \ \ \ \rm{and} \ \ \ \ \ \
A_t(\theta)\ =\ \frac{\sin^4\theta}{(3+\cos^2\theta)^2}
\label{eq2}
\end{equation}

\noindent
and shown as a function of $\cos\theta$ in Fig. \ref{asyfun}.
To note is that both $A_L(\theta)$ and $A_t(\theta)$ are small
in the forward direction so that the asymmetries
are small in the region where the t-channel diagram dominates.\\ 

\vspace*{-2mm}

\noindent
In order to determine the beam polarisation, the rate of the electrons
scattered in a given solid angle $d\Omega$ is measured in one
orientation of the beam and target polarisation vectors 
$(\vec{P}\,{^B},\vec{P}\,{^T})$ and then
with the beam polarisation vector inverted i.e., $(-\vec{P}\,{^B},\vec{P}\,{^T})$.
Here the  polarisation vectors are defined as
$\vec{P}\,{^B} \equiv [P^B_L,P^B_t]$ 
for the beam and $\vec{P}\,{^T} \equiv [P^T_L,P^T_t]$ for
the target.
The longitudinal component,  $P_L$, is in the z-direction and
$P_t$, the transverse component, is perpendicular to that direction.
Thus the two rates which one measures are:
\begin{center}
\[ {\it R}(s,\vec{P}\,{^B},\vec{P}\,{^T})\ \ \ \ \ {\rm {and}} \ \ \ \ \ 
{\it R}(s,-\vec{P}\,{^B},\vec{P}\,{^T}) \,,\]  
\end{center}
normalised to the same integrated luminosity.
From these rates one constructs the asymmetry $A_R$ which is equal to:

\begin{equation}
A_R\equiv \frac{{\it R}(s,\vec{P}\,{^B},\vec{P}\,{^T})
-{\it R}(s,-\vec{P}\,{^B},\vec{P}\,{^T})}{
{\it R}(s,\vec{P}\,{^B},\vec{P}\,{^T})+{\it R}(s,-\vec{P}\,{^B},\vec{P}\,{^T})}
=- P_L^BP_L^TA_L(\theta)-P_t^BP_t^TA_t(\theta)
\cos(2\phi-\phi_B-\phi_T)
\label{eq3}
\end{equation}   
Finally the beam polarisation is extracted from the measured values of $A_R$,
the measured target polarisation and the unpolarised asymmetry functions
 given in Eq. \ref{eq2}.\\

\noindent
The expressions given in Eqs. \ref{mollerf} and \ref{eq3} 
are derived for
the lowest order diagrams for the  
$e^-e^- \to e^- e^-$ process.
The contributions of higher diagrams, up to order 4 in the fine structure 
constant $\alpha$, were in the past investigated 
in  \cite{alpha4} and more recently in \cite{mol-rad}.
The QED corrections to the M{\o}ller asymmetry can be evaluated through 
the BMOLLR code developed by S. Jadach and B. Ward \cite{bmollr}. 
This is an $\mathcal {O}(\alpha)$ exponentiated Monte Carlo generator for 
$e^-e^- \to e^- e^- \, + \,n\gamma$ with any $n$ value.
\subsection{Some general features}
Some obvious features of the M\o ller scattering can be deduced
 from Eq. \ref{mollerf}.\\  

\noindent
a) The cross section is seen to diverge at $\cos\theta$ = $\pm$1.
This is due to the fact that the electron mass was neglected. In a 
rigorous treatment, where $m_e$ is not neglected,
the M{\o}ller scattering formula remains finite
even at $\cos\theta$ = $\pm$1.\\
b) The cross section magnitude decreases as ${\boldmath s}$ increases, 
similar to the one photon annihilation process in $e^+e^-$ annihilation.\\
c) Only if the beam and the target are simultaneously transverse and/or
longitudinal polarised a change in the M{\o}ller scattering 
will be observed.\\
d) In the absence of transverse polarisation the cross section is 
independent of the azimuthal angle $\phi$. This independence 
can also be achieved by integrating over the whole azimuthal
angle $\phi$ provided of course that the experimental setup is $\phi$
independent.\\
e) The asymmetry functions reach their maximum at a  
CM scattering angle of 90$^o$ 
and approach zero in the forward and backward directions.\\

\subsection{Differential cross sections and asymmetries}
\begin{figure}[htbp]
\centerline{\hbox{
\psfig{figure=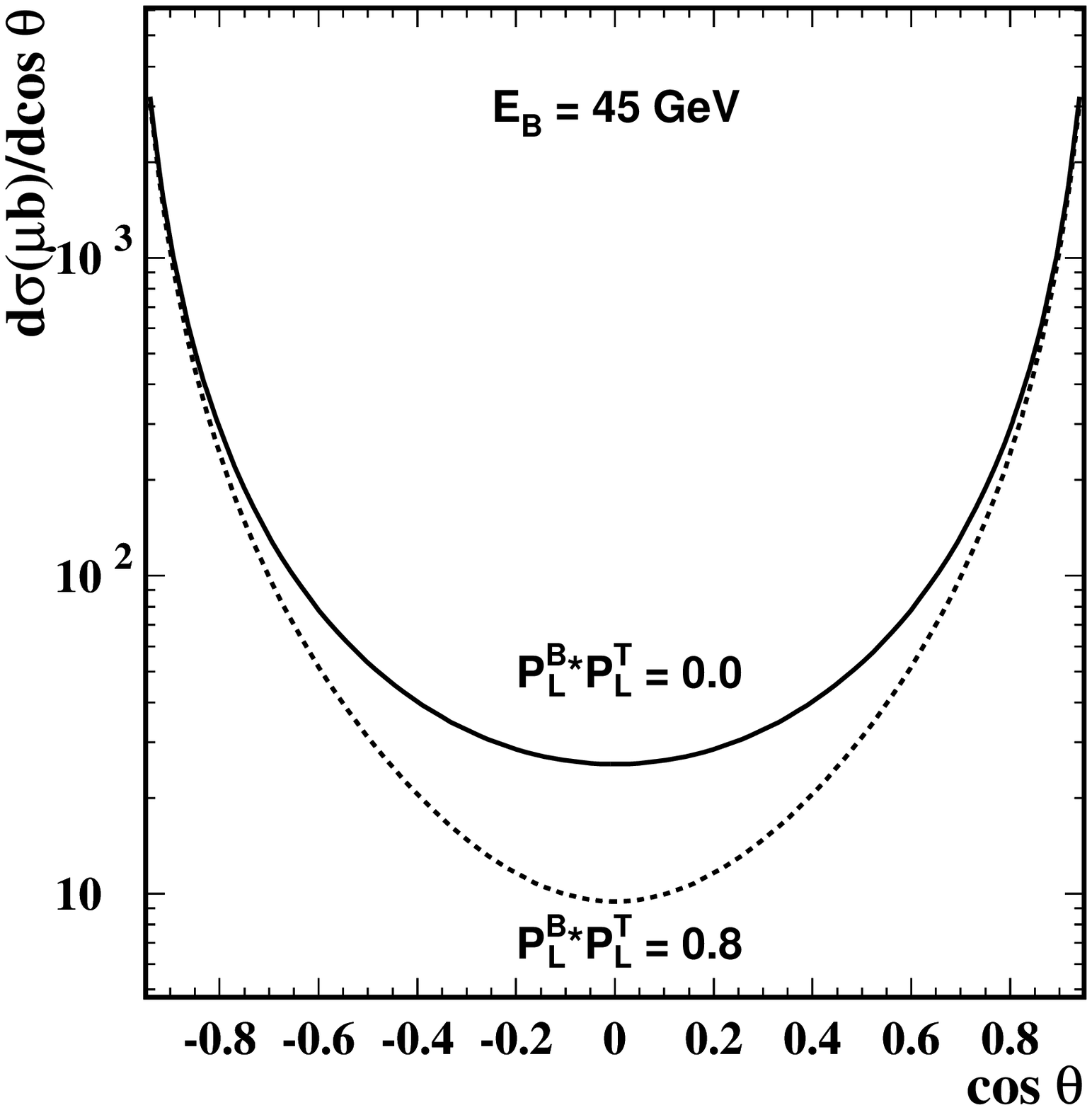,width= 7cm,height= 6.9cm}
\psfig{figure=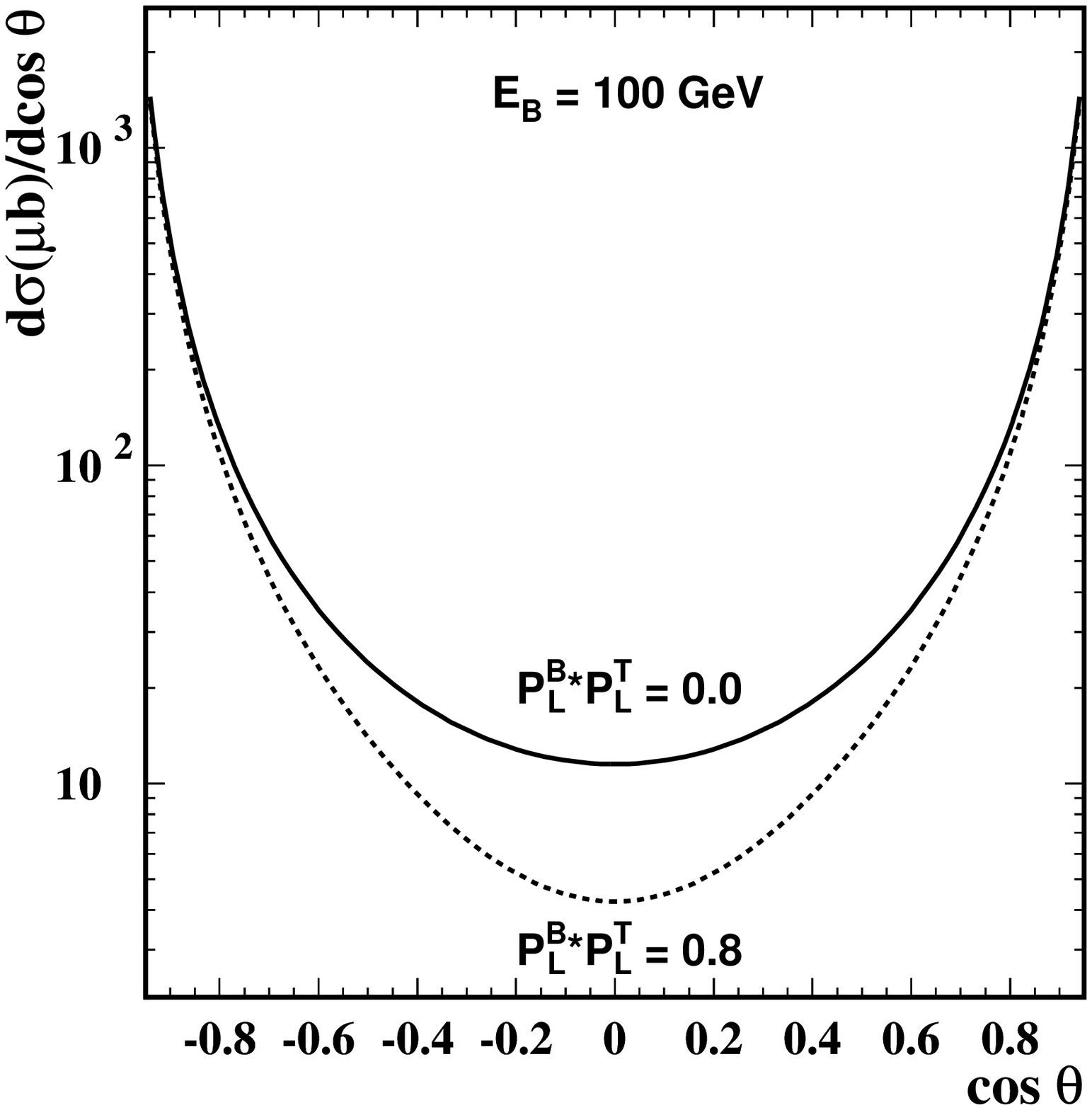,width= 7cm,height= 6.9cm}
}}
\centerline{\hbox{
\psfig{figure=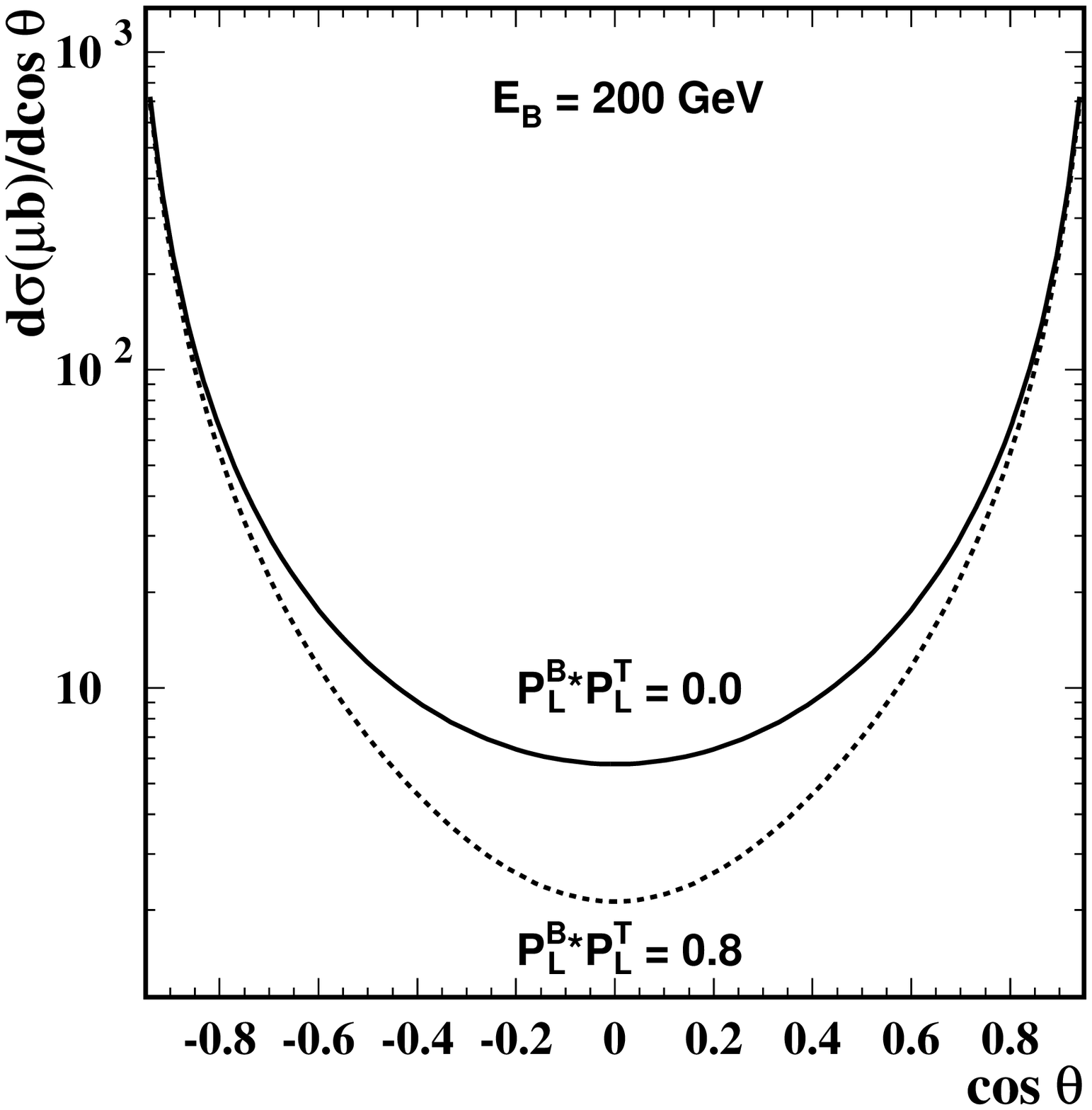,width= 7cm,height= 6.9cm}
\psfig{figure=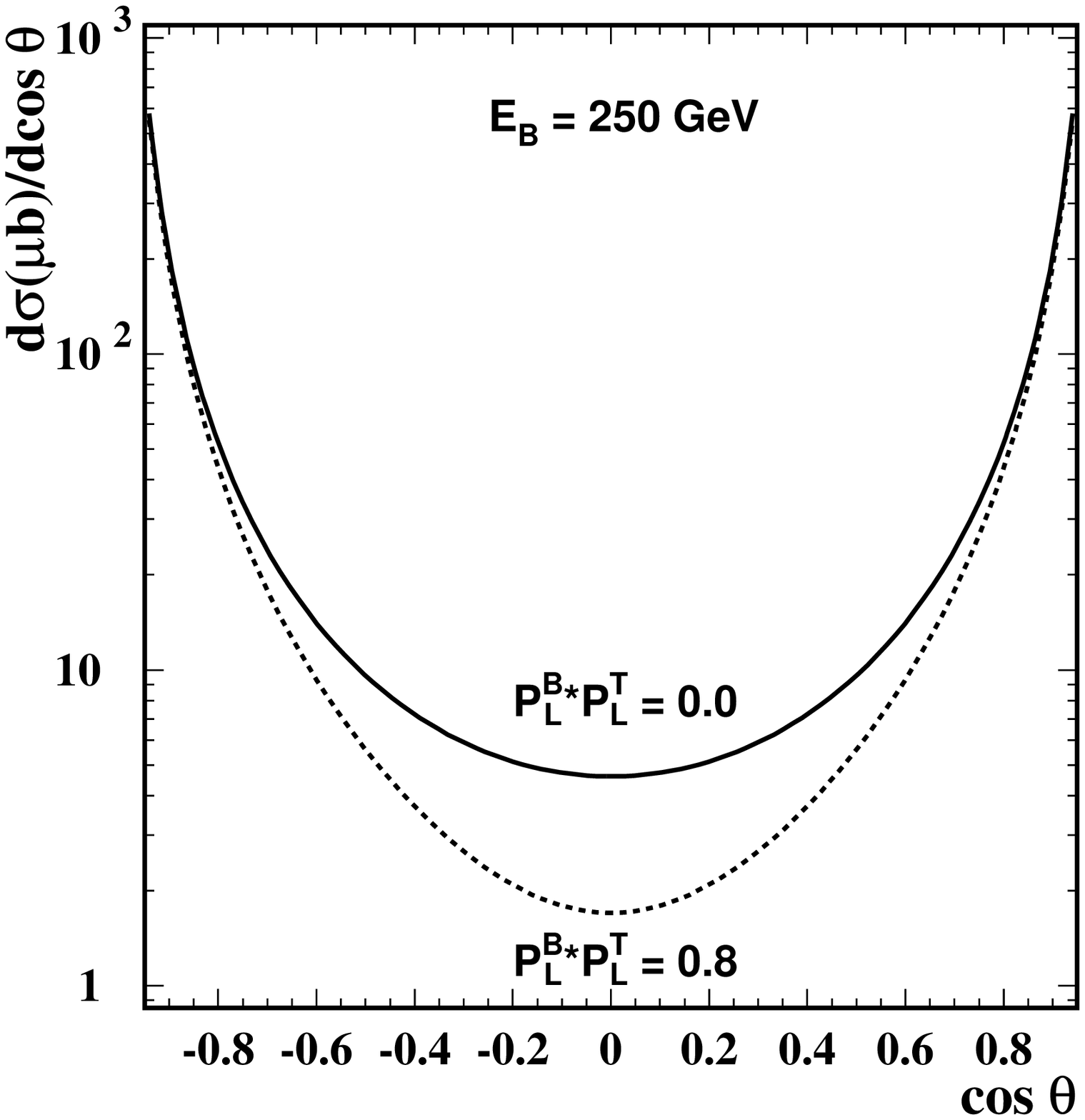,width= 7cm,height= 6.9cm}
}}
\centerline{\hbox{
\psfig{figure=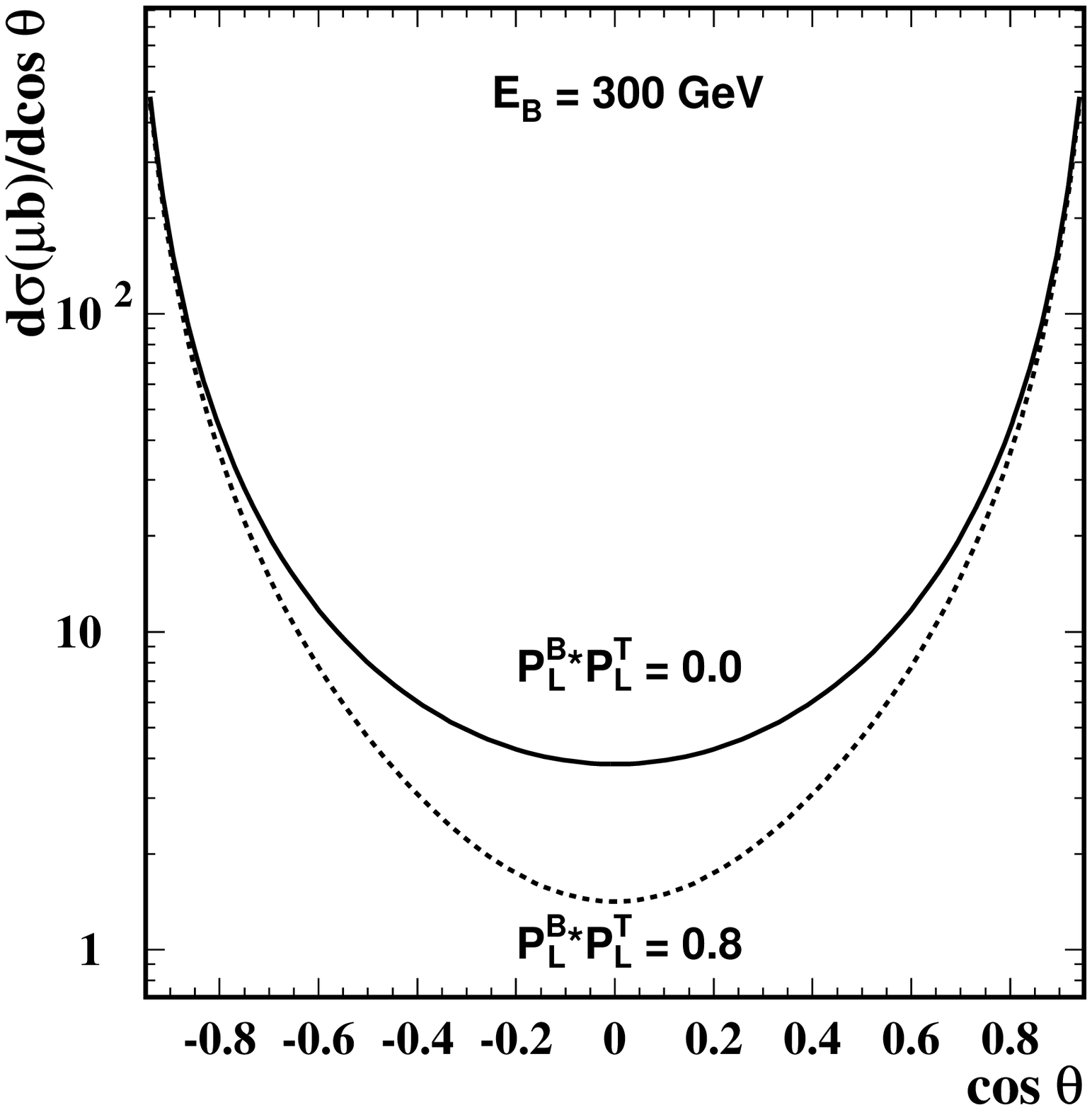,width= 7cm,height= 6.9cm}
\psfig{figure=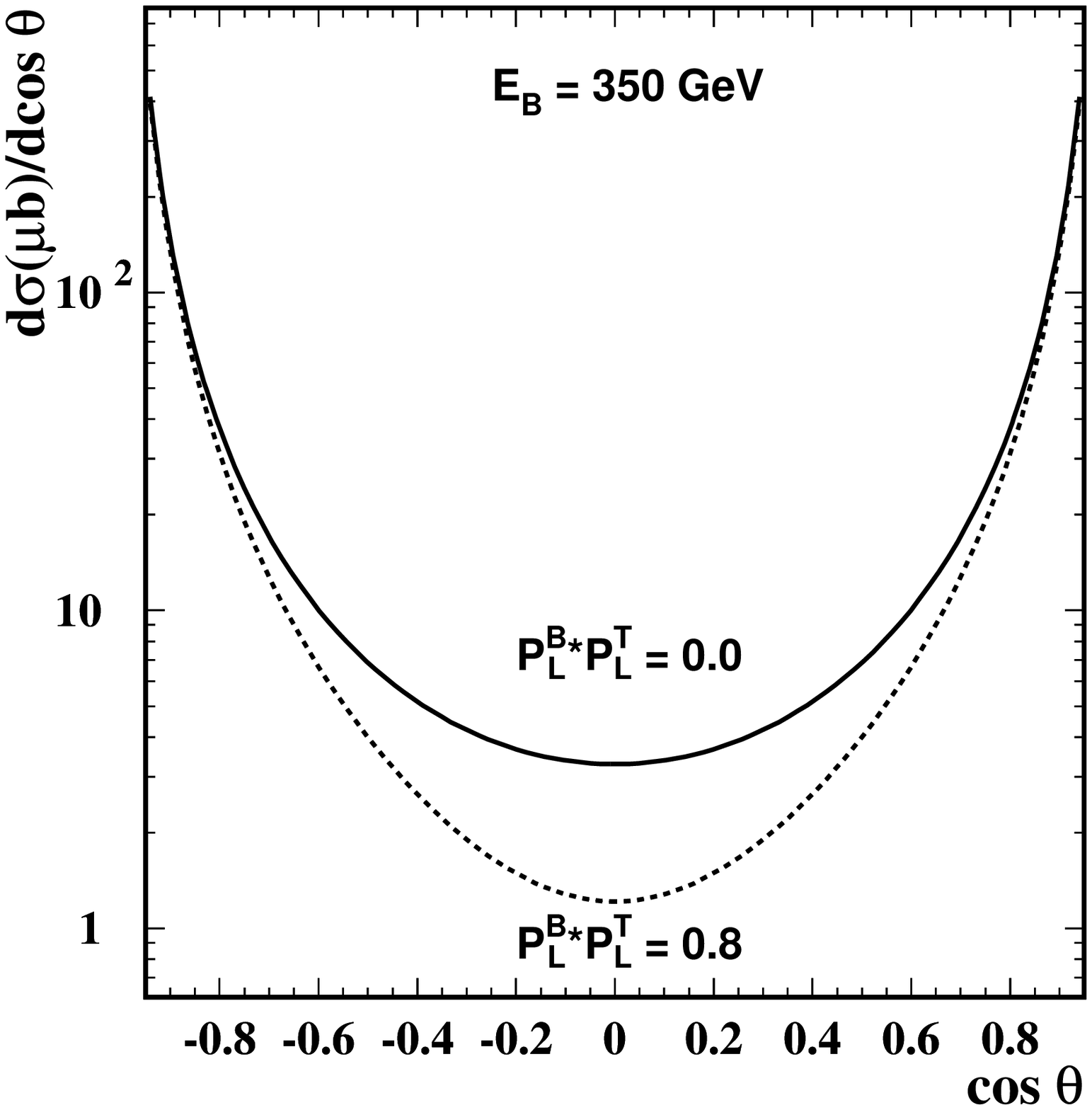,width= 7cm,height= 6.9cm}
}}
\caption{M{\o}ller differential cross section, $d\sigma /d\cos\theta$,
as a function of $\cos\theta$ in the range of ~$|\cos\theta| < 0.9$ 
for several $E_B$ and without and with longitudinal
polarisation.}
\label{moldiff}
\end{figure}

\begin{figure}
\centerline{\hbox{
\psfig{figure=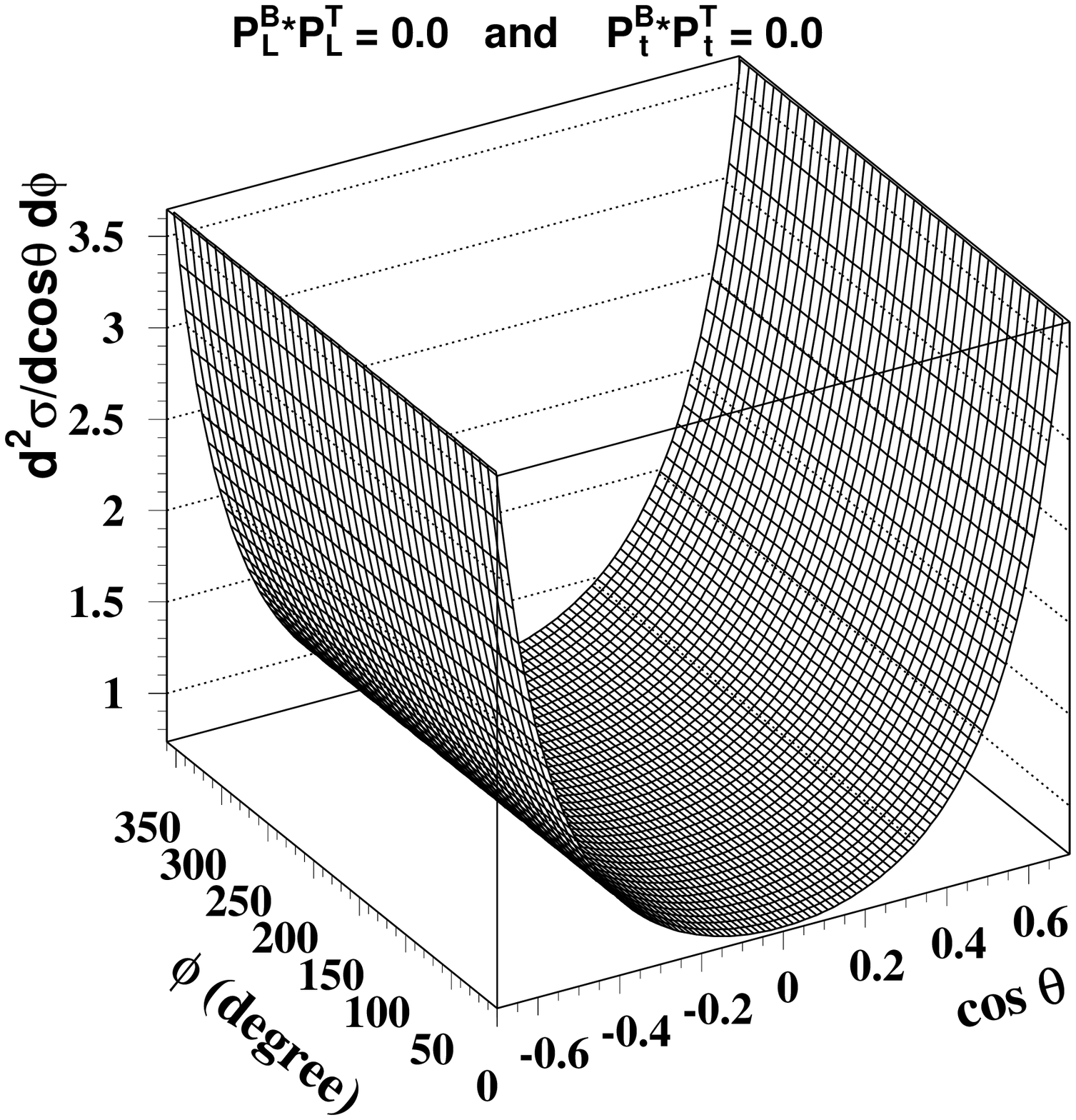,width= 9cm,height=9cm}
\psfig{figure=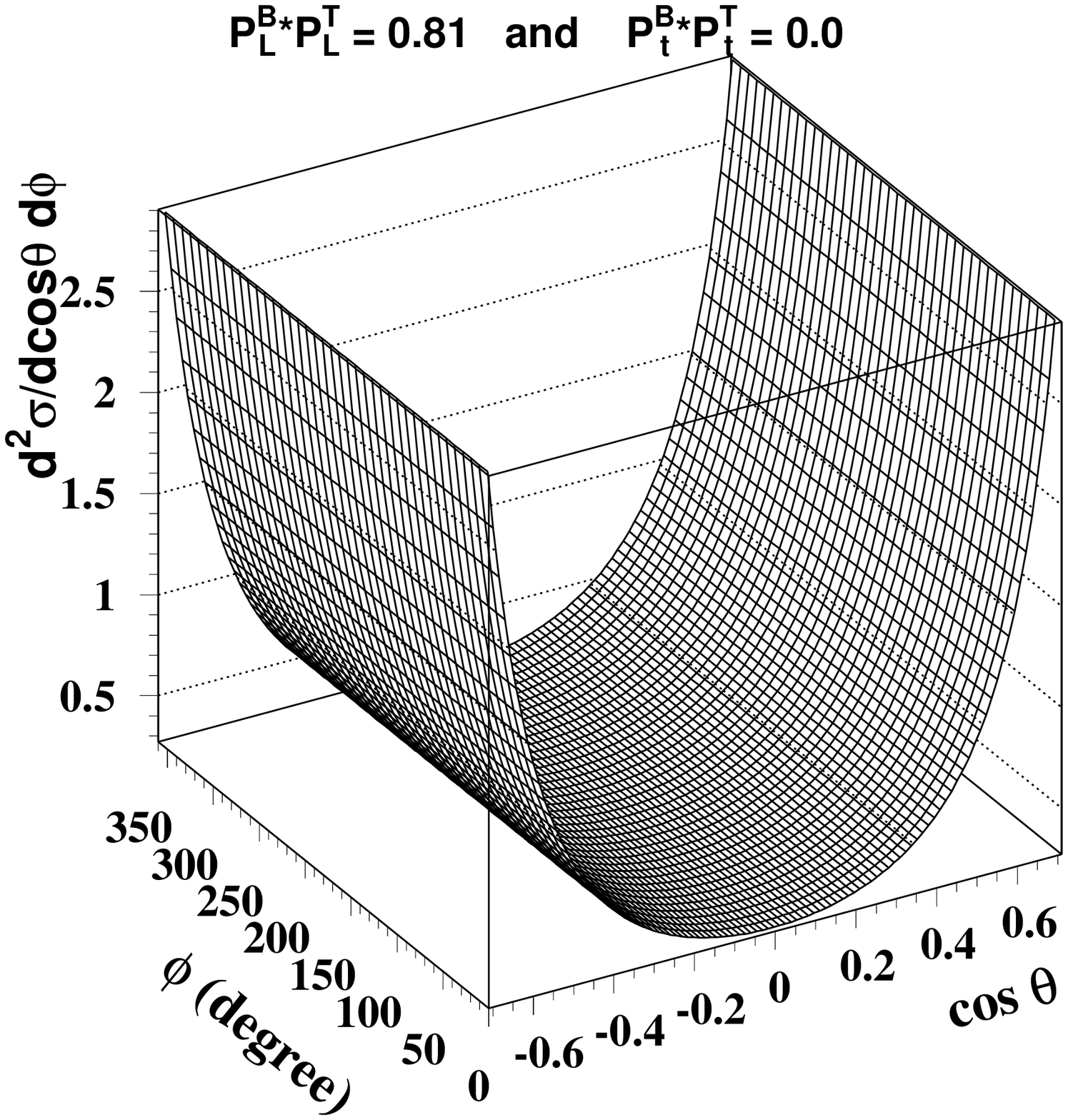,width= 9cm,height=9cm}
}}
\centerline{\hbox{
\psfig{figure=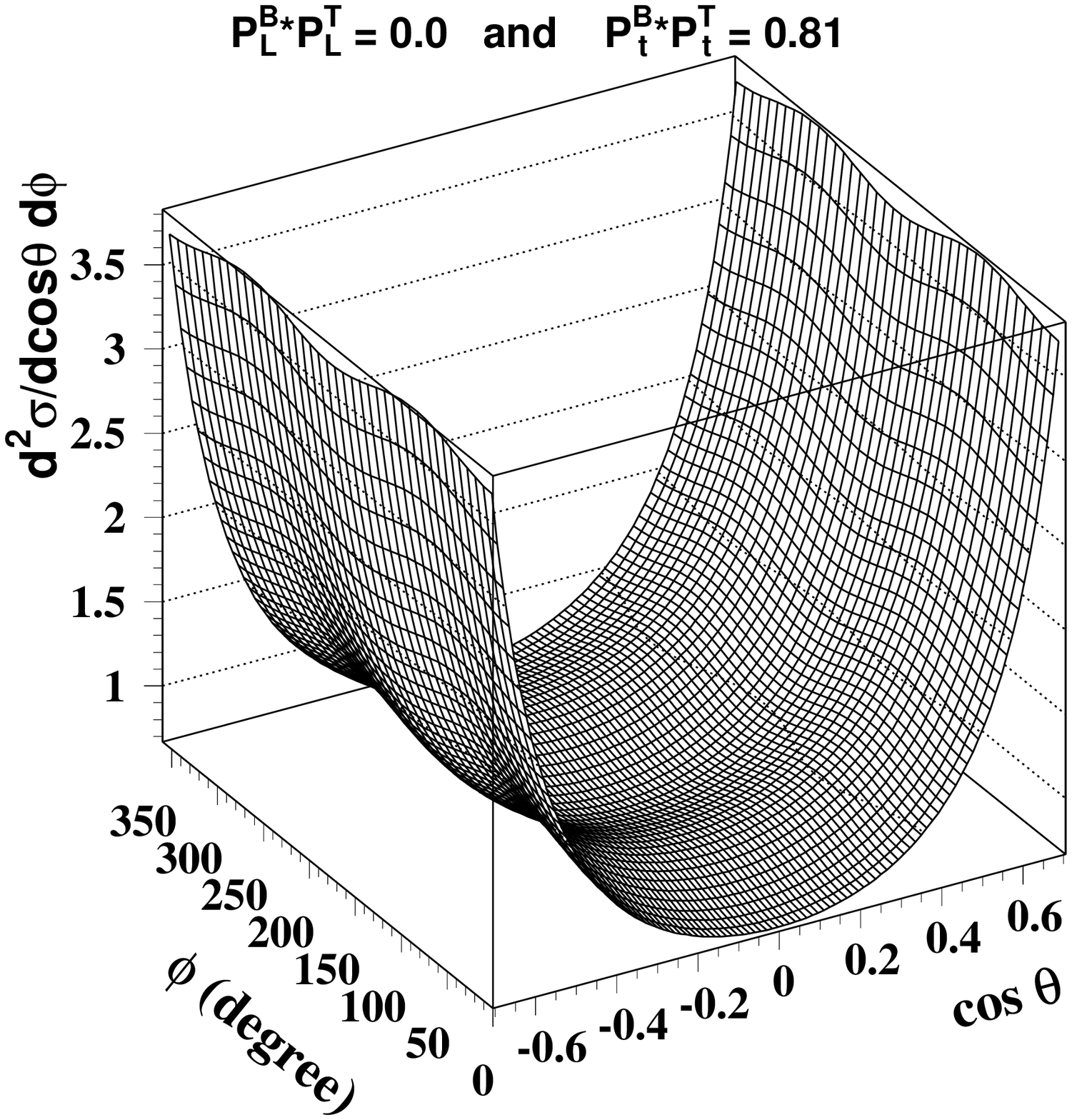,width= 9cm,height=9cm}
\psfig{figure=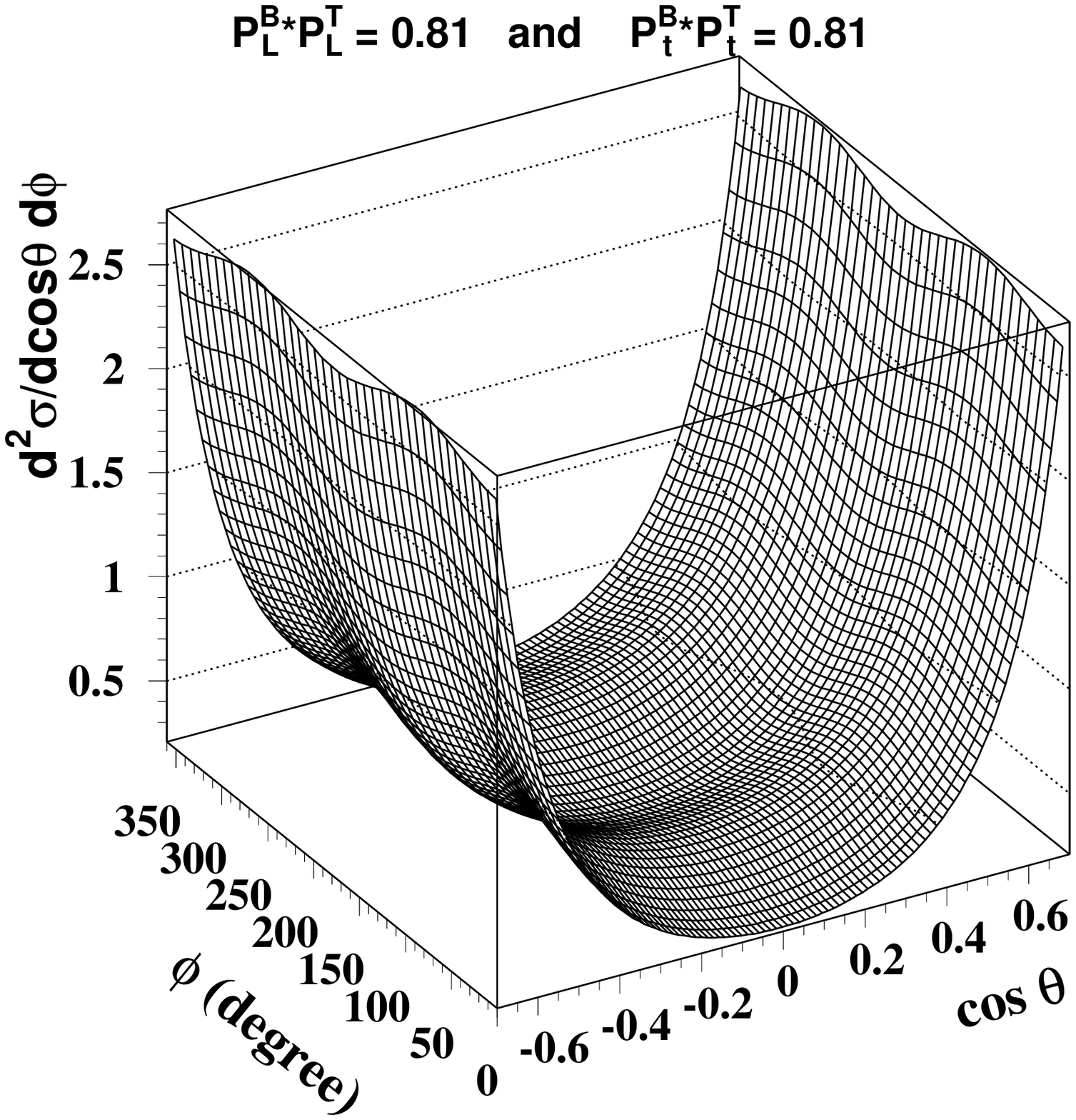,width= 9cm,height=9cm}
}}
\caption{The 2-dimensional plots of 
$d^2\sigma(e^-e^- \to e^-e^-)/(d\cos\theta d\phi)$ in $\mu$b,
at $E_B = 250$ GeV, are
shown for several values of the longitudinal and transversal
 polarisations as indicated in the figures.}
\label{twod}
\end{figure}

\begin{figure}
\centerline{\psfig{figure=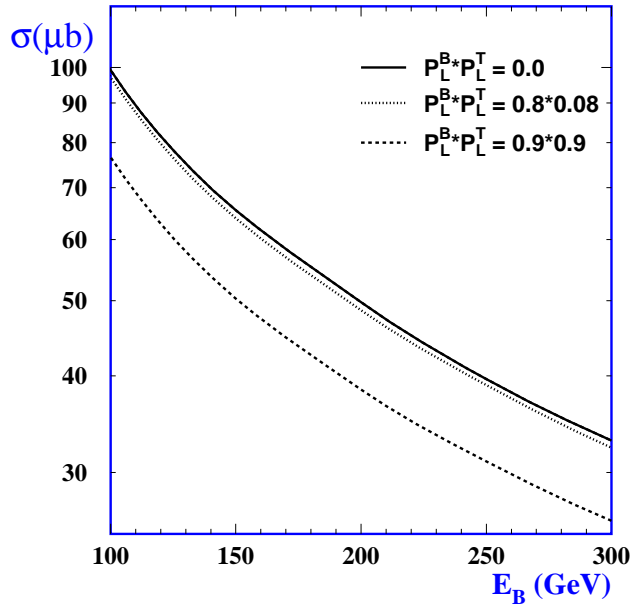,width= 9cm,height=9cm}}
\caption{The total M{\o}ller scattering cross section, integrated over the range
$0^o < \phi < 360^o$ and  $\arrowvert \cos\theta \arrowvert < 0.9$, 
as a function of the beam energy
without and with longitudinal polarisation. The target electron is taken to be
at rest.}
\label{totalx}
\end{figure}

\begin{figure}
\centerline{\hbox{
\psfig{figure=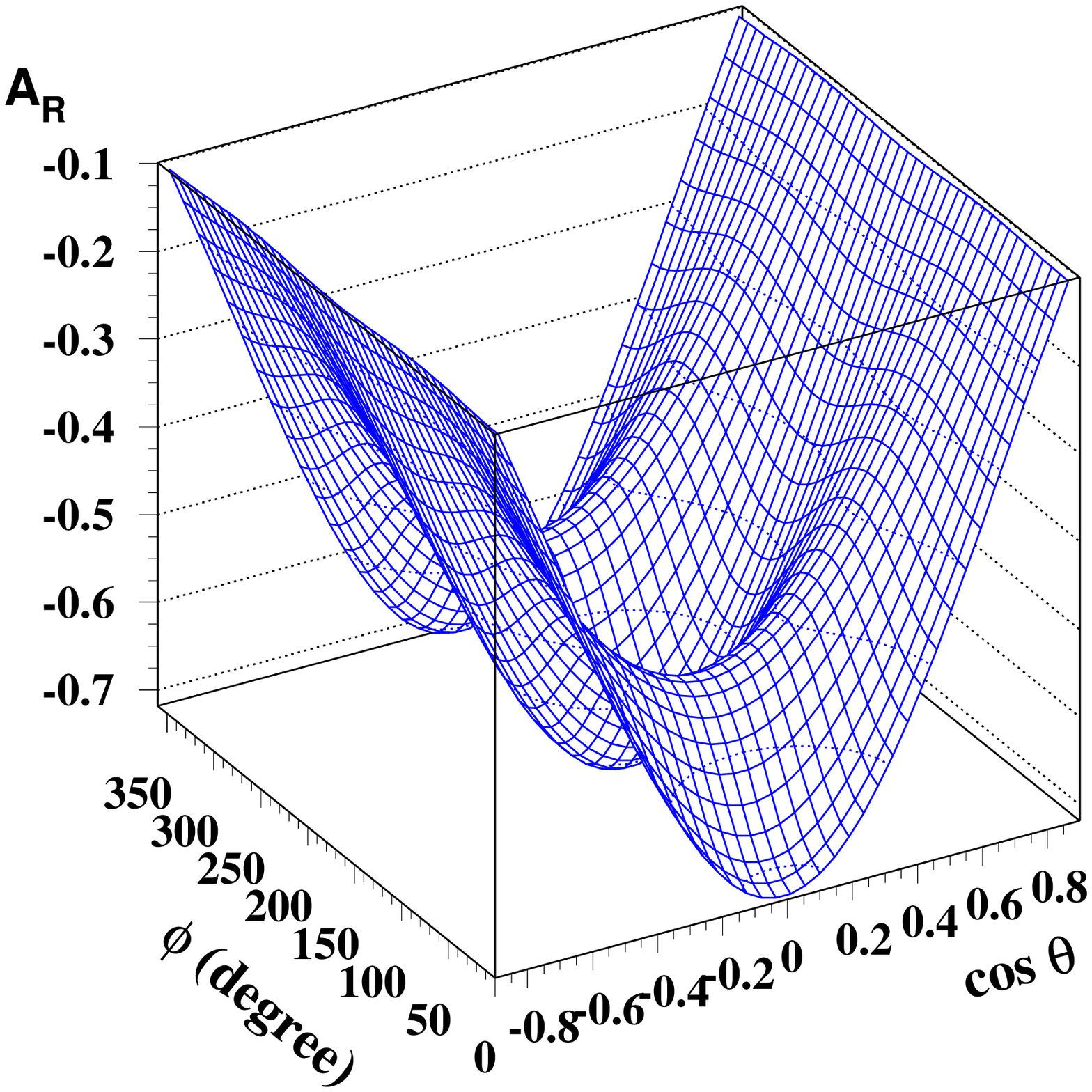,width= 8.7cm,height=8.7cm}
\psfig{figure=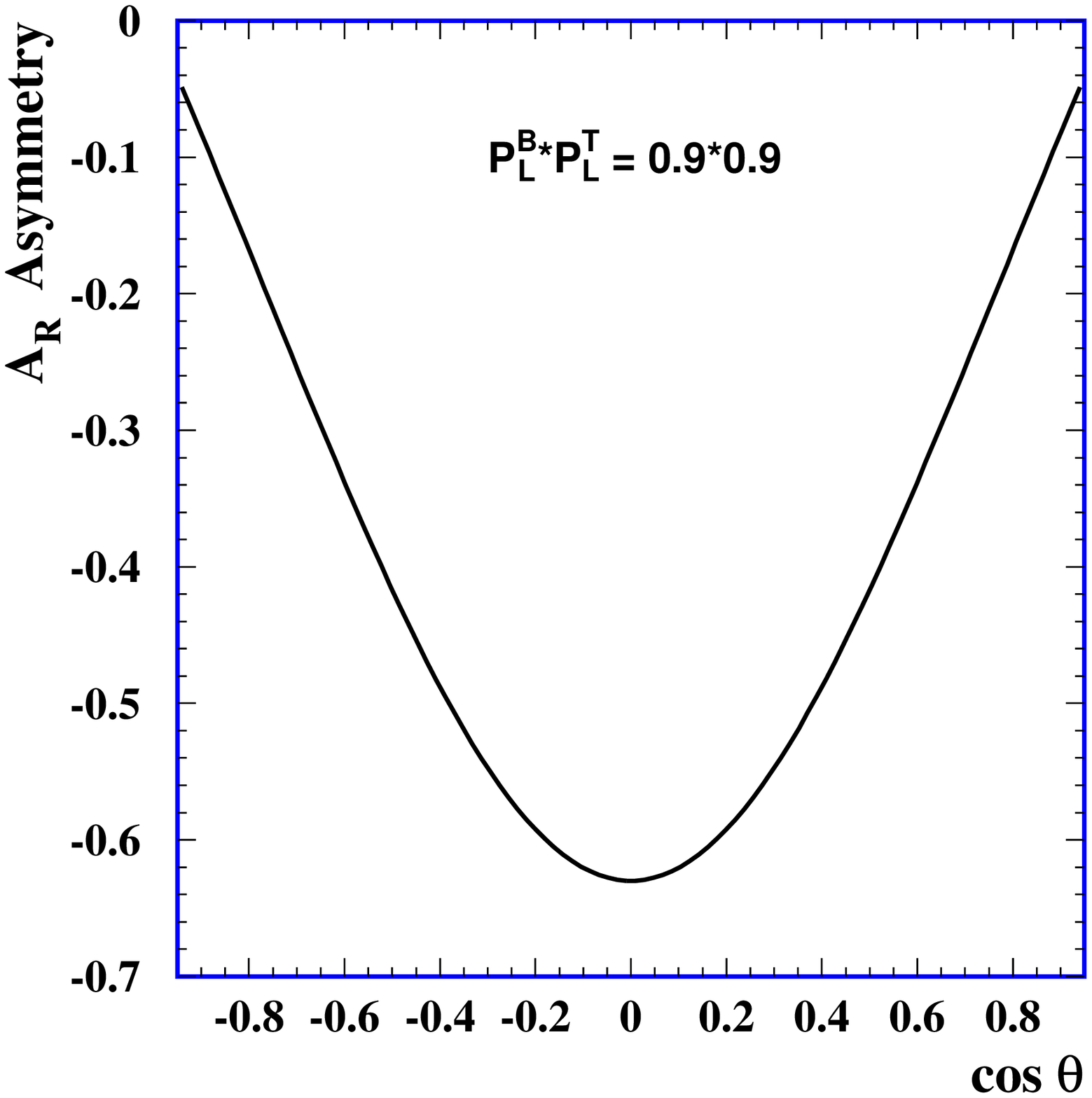,width= 8.3cm,height=8.3cm}
}}
\caption{The M{\o}ller scattering asymmetry $A_R$ in the CM system.
Left, a 2-dimensional plot of $A_R$ as a function of  $\cos\theta$
and  the azimuthal angle $\phi$ 
for $P^B_LP^T_L = 0.81$  and $P^B_tP^T_t = 0.81$.
Right, $A_R$ as a function of  $\cos\theta$, integrated over
$0^o < \phi < 360^o$, the beam and target polarisations values are indicated 
in the figure.} 
\label{asymm}
\end{figure}

The cross sections and asymmetries were studied for a set of $E_B$
values and several beam-target polarisation configurations where we
set everywhere $\phi_B + \phi_T$ to zero.
In order to illustrate the effects of the beam and target polarisations
we assume for the target a polarisation of 90$\%$ realising that in practice
an iron target has a maximum  polarisation of 8$\%$. The beam 
polarisation was taken to be 90$\%$.
In  Fig. \ref{moldiff} we show, after integrating
over the whole azimuthal angle $\phi$, typical CM polar angular distributions
of the M{\o}ller scattering at different beam energies 
scattered by a stationary electron target, without and with
longitudinal polarisation characterised  by their
~$P^B_LP^T_L$ values. 
For  beam electrons of $E_B = 250$ GeV scattered over 
a stationary electron, we show in Fig. \ref{twod}  
the two dimensional 
plots of the M{\o}ller differential cross section 
$d^2\sigma/(d\cos\theta d\phi)$
with and without longitudinal polarisation and without and with 
transverse polarisation.  
In Fig. \ref{totalx} we show the total M{\o}ller cross section
integrated over $\phi$ from $0^o$ to  $360^o$ and over the polar angle
region $\arrowvert \cos\theta \arrowvert < 0.9$ for unpolarised and 
polarised beam and target.
Finally we show in Fig. \ref{asymm}(left) the M{\o}ller scattering asymmetry
$A_R$ as a function of $\cos\theta$ and $\phi$ for a longitudinal polarisation
value of $P^B_LP^T_L = 0.81$ and $P^B_tP^T_t = 0.81$. In the same figure
we also show the asymmetry $A_R$ dependence on  
$\cos\theta$ after integrating over the  
$\phi$ angle.  
\section{M{\o}ller scattering in the laboratory system}
In the collision of two electrons the total centre of mass
energy squared $s$ is written in the Lorentz invariant form, 
and thus valid in any reference system, as:

\begin{equation}
s  = 2m^2_e + 2E_BE_T(1-\beta_B\beta_T\cos\theta_{1,2}) ~,
\label{s_all}
\end{equation}
where $E_B$ and $E_T$ are the beam and target electron energies, 
$\theta_{1,2}$ is the angle between the incident particle's momenta and 
$\beta_{B,T} = {p}_{B,T}/E_{B,T}$ are the velocities of beam and
 target electrons.\\   

\noindent
At TESLA, where the beam energy is planned to be a few hundreds GeV, 
one has  $\beta_B = p_{_B}/E_{_B} \simeq 1$  so that one can neglect  
$m^2_e$ and Eq. \ref{s_all} 
reduces to:
\begin{equation}
   s  =  2E_BE_T(1-\beta_T\cos\theta_{1,2}) ~.
\label{s_tesla}
\end{equation} 

\subsection{Target electrons at rest}

In the approximation that the target electron is in the laboratory
system a free particle at rest
($\beta_T = 0$), the square of 
the centre of mass energy, $ s_0 $, is given by the expression:
 \begin{equation}
   s_0  =  2E_Bm_e =  2p_{_B}m_e ~,
\label{s_0}
\end{equation}
where $p_{_B}$ is the beam momentum and $m_e$ is the electron mass.
The laboratory momentum of the scattered electron , $p_{lab}$, is 
given by :

\begin{equation}
p_{lab} = \gamma_{_{\rm CM}}\sqrt{\left ( E^\ast + p^\ast\beta_{_{\rm CM}}\cos\theta \right )^2 -
 \frac{m_e^2}{\gamma_{_{\rm CM}}^2}} ~,
\label{p_labgeneral}
\end{equation}
where $ ~p^\ast, ~E^\ast$ are the momentum and energy of 
the incident electron in the CM system
and $ \gamma_{_{\rm CM}} = 1/\sqrt{1- \beta^2_{_{\rm CM}}}$.
The relation between the laboratory scattering angle $\theta_{lab}$
and the CM scattering angle $\theta$ is given by:
\begin{equation}
\tan\theta_{lab} =
\frac{1}{\gamma_{_{\rm CM}}} \times \frac{\sin \theta}{\rho + cos \theta} ~,
\label{angle_transform}
\end{equation}

\noindent
\rm{where} \ 
$\rho = \beta_{_{\rm CM}} / \beta^\ast$ is the ratio of the velocity of the 
center of mass system and the velocity of the electron in the CM system.  
For elastic scattering of a beam electron of ~$E_B $= 250 GeV on an
electron at rest one has  
$m_e / E_{_B} =2\times10^{-6} << 1$ 
so that
Eqs. \ref{p_labgeneral} and \ref{angle_transform} can be 
reduced to a simpler form. \\

\noindent
Using the relation between $ \beta_{_{\rm CM}}$ and 
~$\gamma_{_{\rm CM}}$ and remembering that 
\[\beta_{_{\rm CM}}\ = p_{_B}/(E_B +m_e)\]
one obtains:
\begin {equation}
\gamma_{_{\rm CM}} =\sqrt{ (E_B +m_e)/2m_e} ~.
\label{lorentz} 
\end{equation}
In the CM system  
of the M\o ller scattering the momentum and energy of the incident electron
are expressed by:
\begin{eqnarray} 
 p^\ast   =  m_e\sqrt{\frac{E_B^2-m_e^2}{2m_e^2 + 2E_Bm_e}} & = & \sqrt{\frac{m_e (E_B - m_e)}{2}} \\
 E^\ast   =  m_e \frac{E_B+m_e}{\sqrt{2m_e^2 + 2E_Bm_e}} & = & \sqrt{\frac{m_e (E_B + m_e)}{2}} ~\cdot
\end{eqnarray} 
Therefore 
\begin {equation} 
 \beta^\ast = \frac{E_B-m_e}{E_B+m_e} \ \ \ {\rm and} \ \ \ \rho = \frac{p_{_B}}{E_B-m_e} ~. 
\label{beta-rho}
\end{equation}
Here $\beta^\ast$ is the velocity of 
the incident electron calculated in the CM system.

\noindent
Using Eqs. \ref{s_0}, \ref{lorentz} and \ref{beta-rho}
we rewrite Eq. \ref{p_labgeneral} to be:

\begin{equation}
p_{lab} = \frac{p_{_B}}{\sqrt{s_0}} \times \frac{\sqrt{s_0}}{2}(1 + \cos\theta)
 = \frac{p_{_B}}{2}(1 + \cos\theta) ~\cdot
\label{p_lab}
\end{equation}
From this last equation follows that the momentum of the scattered electron 
does not depend on the CM total energy, but 
only on the beam energy and the CM scattering angle.
From  Eq. \ref{angle_transform} one obtains, in a few 
simple steps, the expression: 
\begin{equation}
\tan^2 \theta_{lab} = \frac{2m_e}{E_B +m_e} \times \frac{1 - \cos\theta}
{1 + \cos\theta} ~\cdot
\label{exact}
\end{equation}

\noindent
Finally in the small angle approximation, where 
$\tan^2\theta_{lab} \simeq \theta^2_{lab}$, one obtains from
Eqs. \ref{p_lab} and \ref{exact} that 
 \begin{equation}
 \theta^2_{lab}  = 2m_e \left (\frac{1}{p_{lab}}- \frac{1}{p_{_B}}\right) ~\cdot
\label{theta_0}
\end{equation}

\begin{figure}
\centerline{\psfig{figure=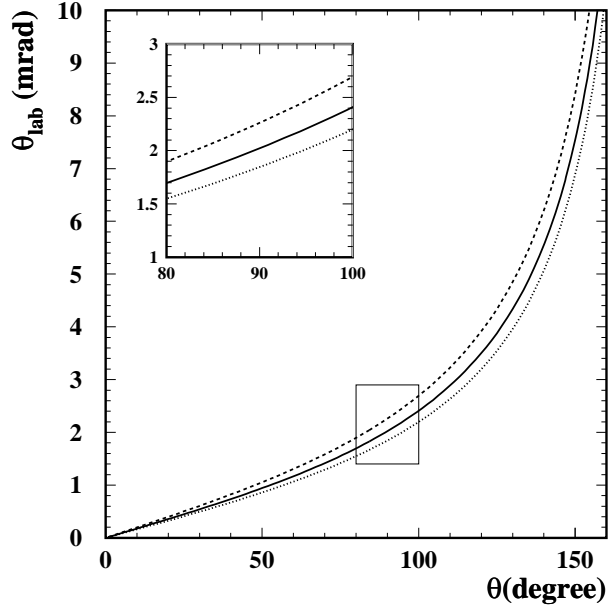,width= 9cm,height= 9cm}}
\caption{The M{\o}ller scattering angle in the laboratory
system, $\theta_{lab}$, as a function of $\theta$, the CM angle.
The target electron is at rest.
The solid line is drawn for $E_B = 250$ GeV , the dashed one 
corresponds to $E_B = 200$ GeV and the dotted line is for $E_B = 300$ GeV  .
The insert magnifies the region where the scattering asymmetry 
is at or near its maximum.}
\label{theta_correl}
\end{figure}

\noindent
The single-arm M{\o}ller polarimetry is based on Eq. \ref{theta_0}
which provides the identification of the elastic $e^-e^-$ scattering
through the relation, in the laboratory system, between $\theta_{lab}$ and $p_{lab}$.
In Fig. \ref{theta_correl} we show the relation between the
centre of mass polar angle $\theta$ and the angle of scattered electron 
$\theta_{lab}$, for three different $E_B$ values.\\

\noindent
{\it The analysing  power} of the M{\o}ller polarimeter 
is proportional to the product of the unpolarized cross
section and the square of the asymmetry \cite{swartz1}. 
The optimal scattering angle for polarimetry is thus the one that maximizes the analysing 
power of the method. 
For longitudinal polarisation
measurements, {\it the analysing  power} is maximum at ~$\theta = 90^o$.
In  Fig. \ref{theta_correl} the  region of $\pm10^o$ around   
this optimal $\theta$ value is magnified.  
For $\theta = 90^o$ one has  for any $E_B$ the 
following  M\o ller scattering relevant quantities:

\begin{center}
$ A_L = \frac{7}{9}$; \ \ \ \ \ $A_t=\frac{1}{9}$  \ \ and \ \ 
$ A_R = -P^B_LP^T_L \times \frac{7}{9},$  
\end{center}

\noindent
so that for $E_B = 250$ GeV one has:\\ 

\noindent
$ {d\sigma}/{d\cos\theta}|_{\cos\theta=0} = 4.55 (1 - 7/9 \times P^B_LP^T_L) ~\mu $barn ~, \ \ \
$p_{lab} = 125$~GeV ~ and\ \   $\theta_{lab} = 2$~mrad.   \\

\noindent
These variables are listed in Table \ref{mol-param} 
for several beam-target polarisation configurations.  

\subsection{Target electrons with non-zero momentum}  

In this subsection we evaluate the effects of the non-zero
momenta of the target electrons on the quantities relevant to the M\o ller
polarimetry. 
The target electrons are in fact not free particles
at rest but are bound to 
atomic shells which in the case of Fe atoms, move with a momentum 
in the range of 
$0 < p_T < 200$ keV \cite{swartz2}.
The kinematic effects of these non-zero momentum
target electrons are similar to those produced by the
initial state radiation, namely the $e^-e^-$ CM energy is modified.\\

\noindent
The detailed kinematics of the scattering of an energetic electron
from a bound state electron moving with momentum $p_{_T}$ is discussed in 
\cite{Levchuk}. To leading order, the CM energy is given by :
\begin{equation}
s = s_0 \left ( 1 - \frac{\vec{p}_{_T} \cdot {\vec n}}{m_e} \right) ~,
\label{s_1}
\end{equation}
where $\vec{p}_{_T}$ is the momentum of the target particle and 
${\vec n}= \vec{p}_{_B}/p_{_B}$ is the unitary vector pointing in the  
direction of the beam particle momentum.
The presence of non-zero momentum target electron does not modify 
Eq. \ref{p_lab} but Eq. \ref{theta_0} is changed  
producing the Levchuk effect namely, the line image 
in $\theta_{lab} - 1/p_{lab}$ space is broader. \\

\noindent
Taking into account the corrected CM energy given by Eq. \ref{s_1}, one
has to modify Eq. \ref{theta_0} to read:
\begin{equation}
\theta^2_{lab}  = 2m_e \left (\frac{1}{p_{lab}}- \frac{1}{p_{_B}}\right)
\left ( 1 \pm \frac{p_{_T}}{m_e} \right) ~\cdot
\label{theta_1}
\end{equation}

\noindent
The laboratory scattering angle is smeared by the square root of the 
target momentum dependent factor  $1 \pm p_{_T} /m_e$,
the same factor which modifies the CM energy.\\

\noindent
In the  experiments which operated during the last decade the 
targets used 
for the M\o ller polarimeters
have been Fe alloy or pure Fe foils
\cite{mol-mit,swartz2,mol-jlab}. In these kind
of materials the K- and L-shell electrons are unpolarised
(with mean momentum of 90 keV and 30 keV)    
and the polarised ones reside in the M- and N-shells (with  10 keV and 2 keV
average momenta). Only two electrons from the M-shell carry 
the Fe magnetization, out 
of a total of 26, yielding a maximum target polarisation of 8$\%$.\\   

\begin{figure}
\centerline{\psfig{figure=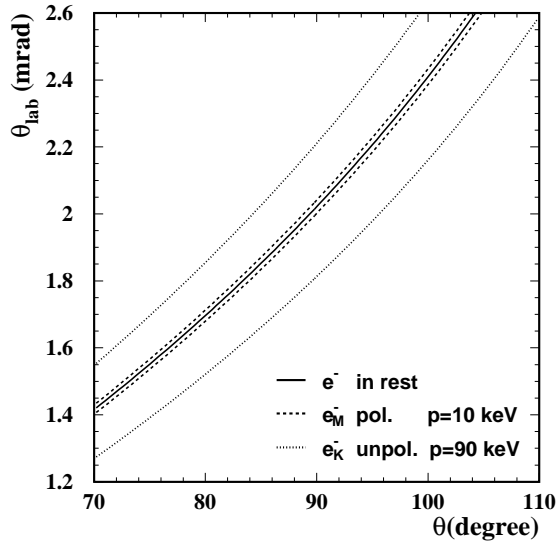,width= 8cm,height=8cm}}
\caption{The M{\o}ller scattering angle, $\theta_{lab}$, in the laboratory
system as a function of $\theta$, the centre of 
mass angle, for $E_B = 250$~GeV.
The solid line is for the target electrons at rest and other two lines
are for electrons from M and K atomic shells having a momentum 
of 10 and 90 keV.}
\label{theta_correl2}
\end{figure}

\noindent
For the polarised target electrons the smearing factor is small 
but for unpolarised electrons it can achieve  values around $20\%$. 
\noindent
In Fig. \ref{theta_correl2} we show the dependence of the scattering angle, 
$\theta_{lab}$, on the CM angle $\theta$ for the case of
target electrons at rest (solid line) and for moving target electrons 
from the 
M and K shells (dashed and dotted lines).\\      

\begin{figure}
\centerline{\psfig{figure=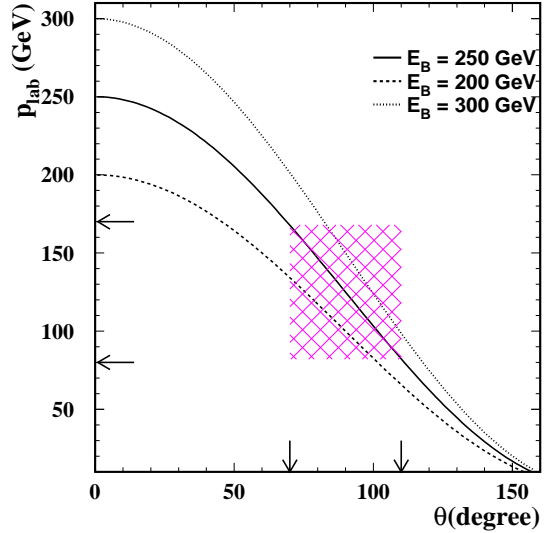,width= 8cm,height=8cm}}
\caption{The M{\o}ller scattered electron momentum 
p$_{lab}$ in the laboratory
system as a function of $\theta$, the centre of mass angle, for various 
beam energy values. The optimal working region is shown 
by the hatched area.}
\label{p_correl}
\end{figure}

\noindent
In Fig. \ref{p_correl} we plot the momentum of elastic scattered M{\o}ller
electrons as a function of $\theta$, the centre of mass angle, for 
$E_B =250$ GeV.
As mentioned earlier, the momentum does not depend on $ s $, 
the total CM 
energy squared
(see Eq. \ref{p_lab}) , therefore it is not modified even when 
the Fermi motion of the target electrons are accounted for.
This means that the lines shown in   
Fig. \ref{p_correl} are not affected by the 
Fermi motion of the electrons and have not to be split for M and K
atomic shells as is the case in Fig. \ref{theta_correl2}.\\ 

\begin{figure}[ht]
\centerline{\psfig{figure=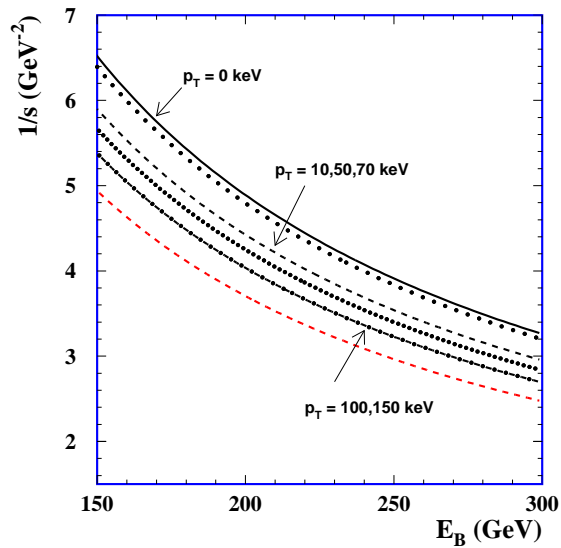,width=8cm,height=8cm}}
\caption{The variation of  $1/s$ with beam energy
for several values of the target electron momentum $p_{T}$.}
\label{factor}
\end{figure}

\noindent
In the case where the electron has a non-zero momentum, the CM energy
squared $s$
is given by Eq. \ref{s_tesla} using the appropriate $\beta_T$ value.
(Note that for target electron at rest $\beta_T = 0$ and one obtains the CM energy squared
$s_0$).    
In order to evaluate the
effect of the non-zero momentum target electrons we
list in 
Table \ref{mol-target} the values for $E^{Kinetic}_T,\ E^{Total}_T$ 
and $\beta_T$  for several value of $P_T$ and in Fig. \ref{factor}
we plot $1/s = 1/[2E_BE_T(1-\beta_B\beta_T \cos\theta_{1,2})]$  as a  function of $E_B$
for various  $p_T$ values setting $\beta_B = 1$ and $\cos\theta_{1,2} = -1$.
From Fig. \ref{factor} follows that
for $E_B = 250$ GeV 
the factor  $1/s$, which enters in Eq. \ref{mollerf}, changes by about 28$\%$ 
when the target electron momentum increases from 0 to 150 keV.\\

\noindent
The effect on the asymmetry however is smaller, namely  $10\, - \,15 \%$ and can be estimated 
by using a proper simulation of the elastic M\o ller signal 
(see e.g. reference \cite{swartz2}). Finally the measured 
polarisation value is shifted 
by about $10\%$ pending on the material of the target, 
the resolution and acceptance of the polarimeter and the 
analysis procedure.\\  

\noindent
The correction needed to account for this shift depends 
on the details of polarimeter construction, the beam parameters and analysis
technique as is pointed out in \cite{swartz2}. 
At SLC the effect was found to be large due to the 
low emittance of the beam and the fine resolution of the detector.
The solution to this problem, 
already adopted by E143 at SLC and 
JLAB \cite{mol-e143,mol-jlab} experiments,
is given by adopting the coincidence measurement method using a
double-arm polarimeter.    
These  polarimeters  have larger acceptance and poorer resolution 
so that the Levchuk effect was shown to be very small. \\

\noindent 
A recent study \cite{Levchuk_2arm} on the performance 
of the double-arm polarimeter,
has discovered a new effect arising from the Fermi motion of the atomic electrons in the target.
Namely, that there exists a dependence of the 
measured M{\o}ller asymmetry on the
relative position of the detectors in the double-arm operation mode.
Here one should point out that the final decision and the technical design
for a M\o ller polarimeter and the corresponding simulation 
of the apparatus and its
performance have to be postponed until a final design of 
the collider will be available.\\

\renewcommand{\arraystretch}{1.6}


\section{Beam - Target related features}

A technical drawing of the TESLA beam line section near
the interaction point (IP),   
where a M\o ller polarimeter could in principle be placed, 
can be seen for example in  Fig. 3.7.2 of Ref. [1], page 474.
The final design and exact position of the M\o ller polarimeter 
however does depend on the choice
taken between a single or a double arm device and on the detailed
knowledge on the beam transport and background conditions.  
In these two possibilities the common elements of a M{\o}ller
polarimeter are:\\

\noindent
$\bullet$ A magnetized target made usually of a Fe or Fe-alloy foil  
or if technological possible  a new type of target (see e.g. Ref. \cite{new-targets});\\ 
$\bullet$ A set of magnets to steer the scattered electrons;\\
$\bullet$ A collimator to define the accepted scattering range
in the azimuthal angle $\phi$
and the  
($\theta_{lab}^{min} - \theta_{lab}^{max}$ ) polar angle interval,  
for the elastic scattered electrons.
This is followed by a dipole magnet that selects  electron momenta 
in the desired range of acceptance;\\ 
$\bullet$ Finally the scattered electrons
are detected and registered by an electromagnetic detector. To this
end one can envisage for example a microstrip Silicon detector coupled
to a electromagnetic calorimeter. This will enable to measure 
simultaneously 
the position and energy of the scattered electrons.
 
\subsection{Expected luminosity}

The current design of the TESLA beam consists of the set
of parameters given in Ref. \cite{conceptual} and  
listed here in Table \ref{tesla_beam}.
This information is used here to evaluate  
{\it the effective luminosity}\footnote{In 
the present context of fixed target scattering 
{\it the effective luminosity} is defined as 
the luminosity of the TESLA beam on the target electrons 
(see Eq. \ref{n_events}).}, 
the expected rate of M\o ller scattering events and its  
heating effect on the target due to the 
energy deposited by the electron beam.
To this end we consider here the following target
and electron beam features.
\begin{itemize}
\item {For the  target we take pure iron foil.  
The electron density $\rho_e^{target}$
in such a pure iron target is:
\begin{equation}
\rho_e^{target} = N_A \cdot \rho \cdot \frac{\bar{Z}}{\bar{A}}
\end{equation}   
where $N_A$ is the Avogadro number, $\rho$ is the density of the
material and  $\bar{Z}, \bar{A}$ are the
mean atomic number and atomic mass of the iron foil.
Taking the density of this material to be $\rho$ = 7.87 g/cm$^3$ 
and  $\bar{Z}/\bar{A} = 466$ kg$^{-1}$,
one obtains: \\
 \begin{equation} \hspace*{-0.9cm} \rho_e^{target} = 2.21  \cdot 10^{24}
  {\mathrm electrons/cm^3} . \end{equation}  }
\item{For the planned beam current of $I_{beam}=45.2 ~\mu$A, the 
number of electrons 
which hit the target in a time interval of one second is: \\
\begin{equation} N^{beam}_e = \frac{I_{beam}}{q_e}= 2.82\cdot 10^{14} ~~{\mathrm electrons/sec} .
 \end{equation} }
\end{itemize}
\subsection{Expected counting rates}  

\noindent    
The counting rate, defined as the number of M\o ller 
scattering events/sec, is given by:
\begin{equation}
{\mathrm Rate} =  d \times \rho_e^{target} \times  N_{e}^{beam} 
\times \, \sigma_{mol} = {\cal L}_M \times \sigma_{mol}
\label{n_events}
\end{equation} 
where ${\cal L}_M$ is the so called effective luminosity.
The M\o ller scattering cross section, ~$\sigma_{mol}$,
is the result of the integration of Eq. \ref{mollerf} over the whole
azimuthal angle range and  
over the CM polar angle $\theta$ domain $\arrowvert\cos\theta \arrowvert
 < 0.34$, chosen for our study, that is: 
\begin{equation}
\sigma_{mol} = \int_{0}^{2\pi} d\phi \int_{-0.34}^{0.34} 
\frac{d\sigma}{d\cos\theta}(1-P_L^BP_L^T A_L(\theta)) ~d\cos\theta ~, 
\label{sigma_cut}
\end{equation}
where\\ 
$$\frac{d\sigma}{dcos\theta} = \frac{\alpha^2}{s}\frac{(3+\cos^2\theta)^2}
{\sin^4\theta}\ .$$ \\
Using the relevant values obtained in subsection 4.1, 
we evaluate for
a high beam polarisation of $ 80\%$ and a Fe target polarisation of $8\%$, 
 namely ~$ P_L^BP_L^T = 0.8\times 0.08$, ~the cross section and effective
luminosity to be: \\

\hspace*{1cm} $\sigma_{mol}  = 3.3 \cdot 10^{-30} ~{\mathrm cm^2} $
  ~~~and ~~~$ {\cal L}_M = 6.2 \cdot 10^{35}$ cm$^{-2}$sec$^{-1}$.\\

\noindent
From Eq. \ref{n_events}
the expected rate of M\o ller events is then: \\
\[{\mathrm  Rate}\, = \,20.6 \cdot 10^5 ~~{\mathrm events/sec} ~. \] 
One should point out that the numbers given here are summed over the whole
$\phi$ angle range from 0 to  $2\pi$. In practice, the polarimeter acceptance
covers only a small part of this range, denoted here by $\Delta \phi$. 
Therefore the last quoted rate has to be
scaled down by the factor  $\Delta \phi / 2\pi$.
For a polarimeter acceptance of  
 $\Delta \phi = 2\pi/9$ rad and  $\Delta \theta = 40^o $ the expected rate is
\[ ~~~~~~~~{\mathrm Rate}\, = \,2.28 \cdot 10^5 ~~{\mathrm events/sec} = 228 ~{\mathrm k\!Hz} ~, \] 
which is well within the range of the current capabilities of the data acquisition systems
used for M\o ller polarimeters.
\subsection{Target for precise M{\o}ller polarimetry}

\noindent
The target frequently used in  M{\o}ller polarimeters consisted of thin foils
of  ferromagnetic alloy
 with the composition 49$\%$ Fe(Z=26, A=57.9), 49$\%$ Co(Z=27, A=58.9) and
 2$\%$ Vanadium(Z=23, A=50.9) known under the name of
Vanadium-Permendur alloy.
It is magnetised {\it in-plane} using a pair of Helmholtz coils 
producing a small magnetic field of 0.01 Tesla. The foils are usually  
mounted under an angle of $\sim 20^o$ with respect to the beam.
The disadvantages of a target with {\it in-plane} magnetisation \footnote{In the 
theory of magnetism the 'magnetisation density' is used, which is
 related linearly with the electron polarisation of the 
ferro-magnetic material.} are:  a poor knowledge of the target polarisation
i.e. having an uncertainty of $\sim 1.5\% - 3.0\%$ as reported in 
Refs.\cite{swartz2,mol-e143,mol-mit,mol-e154}; 
the demagnetisation due to the target heating is undetected and the
target polarisation forms a non-zero angle (typically 20$^o$) with the beam direction.\\ 
      
\noindent
In order to achieve a precision measurements of the beam polarisation 
of $1\% $ these limitations have to be overcome given their massive
contributions to the systematic error of such method.
A novel approach has been recently developed and put to use \cite{basel,mol-jlab}.
This new target is made out of a thin {\it pure iron} foils polarised 
{\it out-of-plane} in saturation with a 4 Tesla magnetic field  parallel to the
electron  beam. 
The online measurement of the relative foil polarisation during the
polarimeter operation is carried out  
with a laser beam making use of the polar Kerr effect \cite{basel}.\\

\noindent
This new target design allows to reach an accuracy of 0.5$\%$
on the target polarisation and thus meets the precision
beam polarisation measurements requested by TESLA.  
This precision, in fact, was obtained in the experiment which operated
with an electron beam of 
few $\mu$A having the energies of 1-6 GeV at JLAB (see Ref. \cite{mol-jlab}). 
   
\subsection{The effect of target temperature rising}

\noindent
In this subsection we address the question of the target heating  
due to the energy deposited by the impinging beam and its possible
effect on the polarimeter performance.
To this end we consider a pure Fe foil target with a thickness of $d = 10
\ \mu$m
and an area of about 30 cm$^2$ so that the target material
seen by the beam is $\widetilde{\rho}= d \times
\rho = 7.87 \cdot 10^{-3}$ g/cm$^2$.
The target, which is cooled down to about 110 K, is polarised 
in saturation {\it out of plane}, 
so that when it is placed perpendicular to the
beam direction the projected longitudinal polarisation of the target
electrons is at its maximum\footnote{If the target is polarised 
{\it in plane} it has to be oriented at an angle $\psi$ relative to the beam 
with the result that the longitudinal
polarisation of the target electrons is reduced to 
$P_L^T = P^T \cdot \cos \psi $ and its
effective thickness is increased to $d_{ef\!f} = d / \sin \psi$.}.

\subsubsection{Local heating per bunch}
\noindent
The TESLA beam design envisages a pulse cycle of 5 Hz. 
The duration of each pulse
is $\sim 1$ msec followed by a pause of $\sim 199$ msec.
Each pulse contains 2820 bunches of 1 psec length which succeeded every
337 nsec.\\

\noindent
The ionisation energy loss in the target of one beam electron with an
energy of 250 GeV, is $dE_{ion}/dx = 2.5\cdot 10^6$ eV/g/cm$^2$
(see e.g. Ref \cite{pdg}).
The energy, ~$E$, ~deposited in the target by one bunch is:
\begin{equation}
E = \widetilde{\rho} \ \times \ dE_{ion}/dx \  \times \  N^{bunch}_e ~,
\label{e_heat_b}
\end{equation} 
where $ N^{bunch}_e $ is the number of beam electrons in one bunch
namely, $2 \cdot 10^{10}$. 
From Eq. \ref{e_heat_b} follows that the quantity of energy deposited in the target 
is $4 \cdot 10^{14}$ ~eV/bunch
corresponding to ~$6.4 \cdot 10^{-5}$ J/bunch.
The local temperature rise induced by this energy depends on the beam
spot size at the position of the polarimeter. 
In the present design of the  Beam Delivery System  of TESLA
a possible position for the Compton and M\o ller polarimeters is the 
straight section of the e$^-$ linac, some few hundred 
meters before the IP just outside the quads \cite{tesla-note}.
Within this straight section 
it is advantageous to place the M{\o}ller polarimeter in the region where
the beam profile is the largest having the dimensions of 
about: $\sigma_x = 75 \mu m$ and $\sigma_y= 7.5\ \mu m$.
For the heating calculations we then take the beam spot to be
$\Delta x = 4\times \sigma_x$ and $\Delta y = 4\times \sigma_y$. 
The instantaneous temperature rise $\Delta$T,
of the target area hit by one beam bunch
over that of the liquid nitrogen,
is shown in Table \ref{tab_temp}.  
The $\Delta$T values given in this Table are calculated for a 
pure iron target of 10 $\mu$m thickness
taking the beam spot area to be $\Delta x \times \Delta y $ for several 
beam profile values. \\ In particular for the case 
$\sigma_x = 75\ \mu m$ and $\sigma_y= 7.5\ \mu m$,  $\Delta$T = 200 K.

\noindent
From the values given in Table \ref{tab_temp} it is obvious
that the local temperature will be  above the melting 
point\footnote{ The Iron melting point is  T = 1808 K.} if all the bunches
within one pulse will hit the same small region of the target foil. 
To avoid this situation
one needs to spread out over the target the
individual hitting positions of the train of bunches. Such a  solution is
the subject of the next subsection.


\subsubsection{The target cooling}

In the present design of the next generation of Linear Colliders
with large beam currents of 30 - 45 $\mu$A and high luminosity values
the typical beam sizes are few tens of microns. Therefore when using
for the beam polarisation measurements a M{\o}ller polarimeter, the 
local target heating requires a special care.\\ 

\noindent
A possible solution to this local heating problem can be the
implementation of a rotating disk target cooled down, 
all around its circumference, to about 110 K by
liquid nitrogen. We note in passing that
such a solution for the heating problem
has been applied to a M\o ller polarimeter \cite{jlab}
which operated in an electron
beam of 0.8 to 5.0 GeV with a current of 0.5 to 5.0 $\mu$A 
where the needed polarisation
measurement time was about 20 min.\\  

\noindent
The envisaged target is placed in such a way that its
rotation axis is parallel displaced to the beam direction so that 
successive bunches hit different regions of the target and in fact they
will be distributed over a circle. Here we note that a
rotating target has already been applied in the experiment described
in Ref. \cite{jlab}.   
If the disk rotates with a frequency of about 1000~Hz, around an axis
displaced by 2.7 cm from the beam direction, all the impact points of the 
2820 bunches of a single pulse, will be spaced by about 60 $\mu$m over
an annular zone having a width of 300 $\mu$m.\\  

\noindent
Inasmuch that one can neglect the interference effects of one heated 
point by the others we evaluate, using the well known heat conduction
formulae \cite{stocker}, that each heated spot will
be cooled down within about 2 msec to 0.1 K above the liquid nitrogen
temperature. 
This short time is mainly determined by the shortest distance
to the cooled edge of the target.
The residual heating of about 0.5 K/sec results in a final
temperature rise of 50 K at the end of a 80 sec operation time needed
for a polarisation measurement with a precision of about 1 \% .
Finally the fact that the pause between pulses is even 199 msec
one has a sufficient safety margin to cover possible additional 
minor factors which may affect the cooling time calculations. \\

\begin{figure}[htbp]
\centerline{\psfig{figure=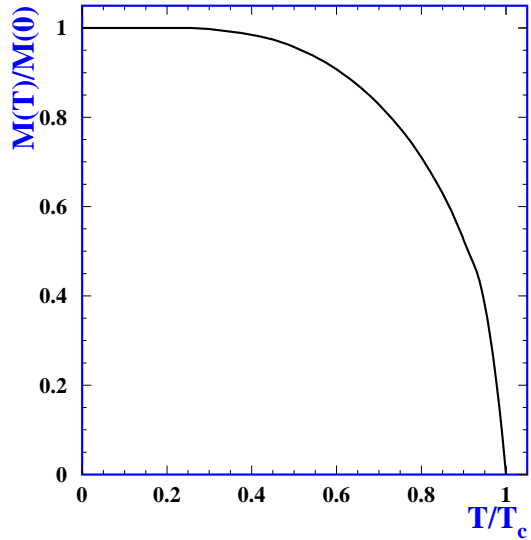,width= 8cm,height=8cm}}
\caption{Saturation magnetisation of Fe as a function of the
temperature scaled by the Curie temperature T$_c$ = 1043 K. 
}
\label{magnet}
\end{figure}

\noindent
The local rise in temperature of about 200 K, causes
a relatively negligible target depolarisation since the working point
at $\sim 110$ K lies in the plateau region 
of the magnetisation curve shown in Fig. \ref{magnet}. 

\noindent
This curve is derived from equation 
\begin{equation}
M(T)\ =\ M(0)\tanh\left (\frac{M(T)}{M(0)}\times\frac{T_c}{T}\right )
\end{equation}
which describes the magnetisation M of the material
as a function of T/T$_c$ where T$_c$ is
the Curie temperature, e.g. 1043 K for Fe
(see e.g. \cite{demag}). For temperatures
above T$_c$ the 
spontaneous magnetisation vanishes. As seen from Fig. \ref{magnet}
at room temperature ($\simeq$ 300 K) or less, the magnetisation
depends weakly on T. Assuming the cooling temperature of the
target to be 110 K, the increase $\Delta$T of the local temperature given in 
Table \ref{tab_temp} corresponds to a change in the magnetisation of 
$\Delta$M = 0.3 $\%$ for a beam profile of $\sigma_x\sigma_y 
= 75 \times 7.5\ \mu$m$^2$.   
Thus the heating of the target does not damage its magnetisation 
and the corresponding polarisation characteristics. 
\section{Measurement of the beam polarisation}
  
The level of the electron beam longitudinal polarisation $P_L^B$ 
is extracted
from the measured $A_R$ asymmetry defined in Eq. \ref{eq3}, 
which is calculated from the results of the two measured 
rates obtained from the M\o ller scattering with   
two different relative orientations of the beam polarisation vector
with respect to that of the target. 
To this end we consider two possible variations for the polarisation 
evaluation:
\begin{itemize}

\item{\bf Integrated polarisation measurement} \\

In this method the number of M\o ller scattering events
is summed over the whole polarimeter acceptance region. 
If we denote by  $N_{+}$ ~and ~$N_{-}$ the recorded scattered events in the
relative parallel
and anti-parallel beam and target polarisation vectors, properly
normalised and corrected for efficiency, then the measured
$A_R$ asymmetry is given by
$(N_{+} - N_{-})/ (N_{+} + N_{-})$. The final beam
polarisation is then derived from this measured
asymmetry and the known target polarisation. 

\item {\bf Differential polarisation measurement}\\

If the polarimeter is equipped with a detector which measures also
the momentum of the M\o ller scattered electrons one is able
not only to measure the total rate of the scattering events but also
their momentum distribution. This allows a more precise
measurement of $P_L^B$ and facilitates a better control of
possible systematic effects. Moreover
it allows to optimise the the polarisation measurement even in those 
cases where the beam is tuned to an energy somewhat different 
from its nominal designed value, e.g. 250 GeV in TESLA. 
In the differential polarisation measurement 
one considers the momentum distribution of the
scattered electrons. For a given set  $N_p$ of $p_{lab}$ momentum bins  the 
numbers $n_{+}^i$ and $n_{-}^i$ of the M{\o}ller electrons are counted.
The asymmetry is calculated for each $x_i$ bin 
and the beam polarisation $P_L^{B,i}$
is at first determined bin by bin. 
The weighted mean of these values provides the final value
of the beam polarisation.   
\end{itemize}   

\subsection{Integrated polarisation measurement}

The numbers of elastic M{\o}ller scattering events are counted for two 
different relative beam-target  polarisation vectors, namely 
$(\vec{P}\,{^B},\vec{P}\,{^T})$ and  $(-\vec{P}\,{^B},\vec{P}\,{^T})$.  
The number of M\o ller events integrated over the  range 
$x_{min}$ to $x_{max}$ 
where $x$ is defined as:
\[ x = 2p_{lab}/E_B  \]
and summed over the measurement time $ T_{+}$
and $T_{-}$ are given by: 
\begin{eqnarray}
N_{+} & = & {\cal L}_{+}T_{+}\int_{x_{min}}^{x_{max}} \varepsilon_{+}(x)
\frac{d\sigma}{dx}(1-P_L^BP_L^T A_L(x)) dx \,,\\
N_{-} & = & {\cal L}_{-}T_{-}\int_{x_{min}}^{x_{max}}  \varepsilon_{-}(x)
\frac{d\sigma}{dx}(1 + P_L^BP_L^T A_L(x))  dx \,,
\label{mollerx}
\end{eqnarray}
where $\varepsilon_{+}(x)$ and $\varepsilon_{-}(x)$ 
describe the efficiencies of the polarimeter as a function of $x$
and 
${\cal L}_{+}$ and ${\cal L}_{-}$ are the luminosity values for
the parallel and anti-parallel beam-target spin states.
The functions
${d\sigma}/{dx}$ and  $A_L(x)$ are the unpolarised  M{\o}ller cross section
and the asymmetry. These are given by the  expressions:
\begin{eqnarray} 
\frac{d\sigma}{dx} & = & \frac{2\pi\alpha^2}{s}\frac{[3 + (x-1)^2]^2}{[1-(x-1)^2]^2}  ~;\\
 A_{L}(x) & = & \frac{[7+(x-1)^2][1-(x-1)^2]}{[3 + (x-1)^2]^2} ~.
\label{asyx}
\end{eqnarray}

\noindent
Eqs. 27 and \ref{asyx} are derived from Eqs. \ref{mollerf}, \ref{eq2} and
the relation \ref{p_lab}
after integration over the whole azimuthal angle range.\\

\noindent
To evaluate the polarimeter performance 
we choose our optimal measurement domain the one shown by the 
stripped area in Fig. \ref{p_correl}, which confines the $x$ range
to the limits   $x_{min}=0.66$  and  $ x_{max}=1.34$ . 
For simplicity we consider the case where the integrated 
luminosities and efficiencies 
for the parallel and anti-parallel polarisations of the 
beam and target electrons are the same 
i.e., ${\cal L}_{+}T_{+}={\cal L}_{-}T_{-}$ and 
$\varepsilon_{+}(x)=\varepsilon_{-}(x)=\varepsilon(x)$.
The experimental measured asymmetry $A_{exp} $ is written as:
\begin{equation}
A_{exp} = \frac{N_{+} - N_{-}}{N_{+} + N_{-}} = P_L^BP_L^T < A_L >  ~,
\label{asy_int}
\end{equation} 
where the mean value $< A_L > $ is given by:
\begin{equation}
< A_L > = \frac{\int \varepsilon(x) \frac{d\sigma}{dx} A_L(x) dx}
{\int \varepsilon(x) \frac{d\sigma}{dx}  dx}  ~.
\end{equation}
Thus the beam polarisation, 
\begin{equation}
P_L^B = \frac{A_{exp}}{P_L^T < A_L >} ~,
\label{P-L,int}
\end{equation}
is proportional to the inverse of the mean longitudinal asymmetry. 
The relative error of the measured beam polarisation $P_L^B$  
is evaluated 
from Eq. \ref{P-L,int} to be:
\begin{equation}
\left (\frac{\Delta P_L^B}{P_L^B}\right )^{2} = \left (\frac{\Delta A_{exp}}{A_{exp}} 
\right )^{2}   + \left (\frac{\Delta P_L^T}{P_L^T} \right )^{2} ~, 
\label{plb_error}
\end{equation}
where the error on the measured asymmetry is :
\begin{equation}
 \Delta A_{exp}^2 = 4 \frac{N_{+} N_{-}}{N ^3} = \frac{1 - A_{exp}^2}{N}=\frac{1}
{{\cal L}T \sigma_t}\left( 1- (P_L^BP_L^T < A_L >)^2 \right ) ~.     
\label{asy-error}
\end{equation}
Here N is the total scattering events number  recorded within
the domain
\mbox{$x_{min} - x_{max}$} i.e.,

\[ N = N_{+} +  N_{-} = {\cal L} \cdot T \cdot \sigma_t  ~,\]
with \[ \sigma_t = \int_{x_{min}}^{x_{max}}\varepsilon(x) \frac{d\sigma}{dx}dx  ~\cdot \] \\ 
By using  Eqs.  \ref{asy-error} and \ref{P-L,int} 
we can rewrite  Eq. \ref{plb_error} as follows:
\begin{equation}
 \left (\frac{\Delta P_L^B}{P_L^B}\right )^{2}  =
  \frac{1}{{\cal L}T \sigma_t}  \frac{1 - (P_L^BP_L^T < A_L >)^2}{(P_L^BP_L^T < A_L >)^2} +
\left (\frac{\Delta P_L^T}{P_L^T} \right )^{2}\ . 
\label{plb2_error}
\end{equation}
Since the term 
$(P_L^BP_L^T < A_L >)^2$ is \ $\leq\ 2.4\cdot 10^{-3}$  
(see Table 4), it is negligible in comparison to 1,
so that one can simplify  Eq. \ref{plb2_error} to:
\begin{equation}
 \left (\frac{\Delta P_L^B}{P_L^B}\right )^{2}   \simeq
\frac{1}{{\cal L}T \sigma_t} \frac{1} {(P_L^BP_L^T < A_L >)^2} 
+ \left (\frac{\Delta P_L^T}{P_L^T} \right )^{2}.
\label{plb3_error}
\end{equation}
Thus the time $t_{Int}$ needed to reach a desired relative
polarisation precision $\Delta P_L^B /P_L^B$ is given by: 
\begin{equation}
\frac{1}{t_{Int}} \simeq {\cal L} \left [ \left ( \frac{\Delta P_L^B}{P_L^B}\right )^2 -  \left (\frac{\Delta P_L^T}{P_L^T} \right )^{2}\right ]
\left (P_L^BP_L^T \right)^2  \sigma_t  <A_L>^2 ,
\end{equation} 
which corresponds to the number of scattering events   
\begin{equation}
N_{Int} = {\cal L} \times t_{Int} \times \sigma_{t} 
\end{equation}
needed to perform the polarisation measurement.
In Table \ref {tab_mol1} we present the 
characteristic unpolarised cross section and the averaged asymmetry values for several 
\mbox{$x_{min} - x_{max}$} regions around $\theta = 90^o$ which may be of 
interest in a M\o ller polarimeter design.
In Table \ref{tab_mol2} we list some of the values concerning the characteristics and
performance of a TESLA M\o ller polarimeter.
Using Eqs. \ref{asy_int} and \ref{asy-error} and the values listed 
in  columns 1 to  3 of the Table \ref{tab_mol2},
we calculated the number of events needed to obtain a  relative 
statistical error of  $ 0.5\%$ for the measured asymmetry $A_{exp}$. 
From this we obtain the 
needed number of events and the corresponding run duration 
for a beam polarisation measurement
with a relative error of about $1\%$ which includes 
an assumed over-all 0.85 $\%$ systematic which includes also
the uncertainty in the target polarisation level. 
These values are shown in the  last two columns
of Table \ref{tab_mol2}. The precision expected from the TESLA
polarimeter in a measurement duration of 80 sec is compared in Table \ref{mol-param2} 
with three existing M\o ller polarimeters attached 
to high energy electron accelerators.

\subsection{Differential polarisation measurement}

The distribution of the measured momentum of the M\o ller scattered
electrons is grouped in several $x_i$ regions. In each region 
the recorded M\o ller scattering events with parallel and anti-parallel
spin configurations are given by: 
\begin{eqnarray}
n_{+}^i & = & {\cal L}_{+}T_{+}\int_{x_i}^{x_{i+1}} \varepsilon_{+}(x)
\frac{d\sigma}{dx}(1-P_L^BP_L^T A_L(x)) dx \\
n_{-}^i & = & {\cal L}_{-}T_{-}\int_{x_i}^{x_{i+1}}  \varepsilon_{-}(x)
\frac{d\sigma}{dx}(1 + P_L^BP_L^T A_L(x))  dx
\label{mollerxd}
\end{eqnarray}
where  ${d\sigma}/{dx}$ and  $A_L$ are the unpolarised  M{\o}ller cross section
and the asymmetry given in Eqs. 27 and \ref{asyx}.\\

\noindent
The experimental asymmetry, in a given $x_i$ bin, is expressed in terms 
of the beam and target polarisations as:
\begin{equation}
A_{exp}^i = \frac{n_{+}^i - n_{-}^i}{n_{+}^i + n_{-}^i} = P_L^{B,i}P_L^T < A_L >^i
\label{asy_diff}
\end{equation}
As previously this formula is for the case where the   
integrated luminosities and efficiencies 
for the parallel and the anti-parallel  electrons polarisation are identical. 
Specifically this means ${\cal L}_{+}T_{+}={\cal L}_{-}T_{-}$
and $\varepsilon_{+} = \varepsilon_{-}$.\\ 

\noindent
For each bin one can then evaluate the beam polarisation as:
\begin{equation}
 P_L^{B,i} = \frac{A_{exp}^i}{P_L^T < A_L >^i} ~\cdot
\end{equation} 
\noindent
The final beam polarisation is obtained as the weighted mean of the measured
polarisations $P_L^{B,i}$, i.e.:  
\begin{equation}
P_L^B = {\sum_{i=1}^{N_p}
\frac{P_L^{B,i}}{\Delta^2 P_L^{B,i}}} \, {\bigg /} \,{\sum_{i=1}^{N_p}
{\frac{1}{\Delta^2 P_L^{B,i}}}}.
\label{P_e_diff}
\end{equation} 
The calculation of the relative error, $\Delta P^B_L/P^B_L$,
follows very closely the one done before for the integrated
polarisation
measurement.   
\noindent
Next we derived an expression for   
the needed time  ~$t_{Dif\!f}$ to achieve a requested 
polarisation measurement 
accuracy of $\Delta P_L^B/P_L^B$, namely:
\begin{equation}
\frac{1}{t_{Dif\!f}} \simeq {\cal L} \left [ \left ( \frac{\Delta P_L^B}{P_L^B}\right )^2
 -\left (\frac{\Delta P_L^T}{P_L^T} \right )^2 \right ]
\left (P_L^BP_L^T \right)^2 \sigma_t  <A_L^2> ~.
\end{equation} 
which translates to the needed scattering events
\begin{equation}
N_{Dif\!f} = {\cal L} \times t_{Dif\!f} \times \sigma_{t} ~~.
\end{equation}

\noindent
Here it should be noted that in the differential method 
a somewhat smaller number
of events is needed to achieve the same precision for the
relative $P_L^B$ measurement.
Another advantage of this method is the fact that it permits a better control
on systematic errors and background contributions.\\

\noindent
Above all the differential
polarisation measurement allows to handle also cases where the linear collider
runs at beam energies slightly away from the nominal designed beam
energy, which in the TESLA case is 250 GeV.
In fact one cannot exclude the necessity to operate the collider at 
beam energies away by several GeV from the nominal 
value due to technical problems or physics needs.
\begin{figure}[htp]
\begin{center}
\psfig{figure=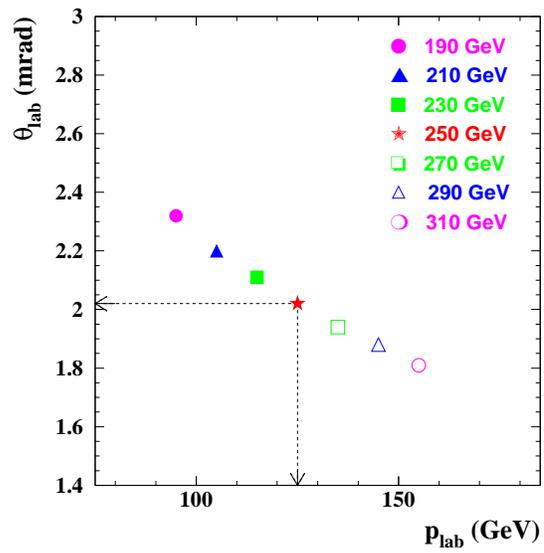,width=8cm,height=8cm}
\end{center}
\caption{The values of $\theta_{lab}$  
and $p_{lab}$ in the laboratory system which correspond to the
90$^o$ centre of mass M\o ller scattering for 
several beam energies around the TESLA nominal value of 250 GeV.}
\label{theta_p}
\end{figure}  
If the beam energy changes the $p_{lab}$ value corresponding 
to $\theta = 90^o$ is
moving and therefore a momentum measurement of the M\o ller scattered electrons
will still allow to utilise those scattered events which yield the maximum 
precision. This is best illustrated in Fig. \ref{theta_p} where
in the plane of $\theta_{lab}$ versus  $p_{lab}$ the position of 
the 90$^o$ centre of mass scattering angle is shown 
for several beam energies around the nominal TESLA value of 250 GeV.  
\vspace{1.5cm}

\noindent
{\large \bf  Acknowledgments}\\

\vspace*{0.2cm}

\noindent
We would like to acknowledge the help and support of 
many members of the TESLA collaboration. In particular our thanks are
due to 
T. Behnke, K. M\"{o}nig and P. Sch\"{u}ler for encouraging us to
study the M\o ller polarimeter option for TESLA. One of us (G.A.) 
would like to thank Profs. T. Hebbeker, T. Lohse and  P. S\"{o}ding
for their very kind hospitality during his stay in the 
Physics Institute
of the Humboldt University, Berlin and in DESY/Zeuthen. Finally this work 
would not have been possible
without the generous financial support extended to him by the DFG during his
stay in Berlin. 

\vspace*{7cm}

\begin{table}[h]
\begin{center}
 \begin{tabular}[h]{|c|c|c|c|c|}
\hline
 $ P_L^B\cdot P_L^T $ & $\theta$ [degrees]& $\theta_{lab}$ [mrad] & $d\sigma/d\cos\theta\ [\mu$barn]  &{\bf A}$_R$   \\ 
\hline   \hline
 $0.0\cdot 0.0$     & 90$^o$   & 2   & 4.6    &   0.0   \\
\hline 
 $0.9\cdot 0.9$     & 90$^o$   & 2   & 1.7    &   0.63   \\
\hline 
 $0.8\cdot 0.08$    & 90$^o$   & 2   & 4.4    &  0.05   \\
\hline 
 \end{tabular}
\caption{\label{mol-param} Some relevant parameters of 
a M{\o}ller polarimeter designed for a 250 GeV electron 
beam operated at centre of mass angle $\theta\ =\ 90^o$
i.e., at its maximum analysing power.}
\end{center}
\end{table}
\begin{table}[h]
\begin{center}
\begin{tabular}[h]{|c|c|c|c|}
\hline
  $ E_T^{Kinetic}$ [keV]   & $E_T^{Total}$ [keV]& ~~$~~\beta_T~~$ & ~~$~~p_T$ [keV] \\ 
\hline   \hline
       ~~0.0             & 511.0   & 0.00  & ~~0.0   \\
\hline 
       ~~0.1            & 511.1   & 0.02  & ~10.0    \\
\hline 
       ~~2.4             & 513.4   & 0.10   & ~50.0    \\
\hline 
       ~~5.0             & 516.0   & 0.14  & ~72.0    \\
\hline 
       ~10.0             & 521.0   & 0.19  & 102.0    \\
\hline 
       ~20.0             & 531.0   & 0.27  & 144.0    \\
\hline 
\end{tabular}
\caption{ \label{mol-target} The electron features of 
an iron target.
The polarised electrons are in the M-shell 
having a momentum in the range $0 < p_T < 75$ ~keV with a
maximum at 10 keV.}
\end{center}
\end{table}

\begin{table}[ht]
\begin{center}
  \begin{tabular}[h]{|r|c|c|}
\hline
Beam Energy ( $E_B$ )   & [GeV] & 250.0  \\ 
 $\gamma=E_B/mc^2$ ( for e ) &   &  $ 4.89 \times 10^{5}$      \\
\hline  
 Horizontal emittance & [m]  & $2.04\times 10^{-11}$ \\  
Vertical emittance  & [m]      & $6.13\times 10^{-14}$     \\
\hline 
 Horizontal normalized emittance & [$\mu$m]  & 10.0     \\
Vertical normalized emittance  & [$\mu$m]    & 0.03     \\
Bunch length at IP &  [mm]    & 0.3 \\
Bunch population   &          & $ 2.0 \times 10^{10}$ \\
Number of bunches  &          &  2820 \\
Bunch separation   & [ns]     & 337 \\
Repetition Rate    & [Hz]     & 5.0 \\
Averaged current   & [$\mu$A] & 45.2 \\
 \hline 
  \end{tabular}
\caption{\label{tesla_beam} TESLA beam characteristics.}
\end{center}
\end{table}

\begin{table}[htb]
\begin{center}
  \begin{tabular}[h]{|c||c|c|c|c|}
\hline
$\Delta x \Delta y [\mu m^2]$ & 400$\times$40 & 320$\times$30 &300$\times$30&240$\times$30 \\ 
\hline
$\Delta$T [$^o$C]& 113& 188 &200 & 250\\
\hline
  \end{tabular}
\caption{\label{tab_temp} 
The instantaneous local 
temperature rise $\Delta$T [$^o$C], of the target within the beam spot 
$\Delta x \Delta y$  during one bunch of $\sim 1$ ps before
the heat is spread over the target.
}
\end{center}
\end{table}

\begin{table}[ht]
\begin{center}
  \begin{tabular}[h]{|c|c|c|c|c|c|}
\hline
$\theta_{CM}\, $[degree]  & $90\, \pm \,10$ & ${\mathbf 90\, \pm \,20}$ & $90\, \pm \,30$ & $90\, \pm \,40$ & $90\, \pm \,50$     \\
\hline
$\theta_{lab}\, $[mrad] & 1.70\, -- \,2.41 & {\bf 1.42\, -- \,2.89} &1.17\, -- \,3.50&0.94\, -- \,4.33& 0.74\, -- \,5.56 \\
\hline
$x_{max}-x_{min}$ &1.17\, -- \,0.83 & {\bf 1.34\, -- \,0.66} &1.5\, -- \,0.5 & 1.64\, -- \,0.36  & 1.76\, -- \,0.26   \\ 
\hline   \hline
$ \sigma_t[2\pi\alpha^2/s]$    &  3.336      &  {\bf 6.831}    &  11.667    &  18.624    & 32.404    \\
\hline 
 $ <A_L> $                 &  0.765     &  {\bf 0.730}   & ~0.668    &  ~0.583   & ~0.459    \\
\hline
 $ <A_L^2> $               & 0.585      &  {\bf 0.534}   & ~0.453    &  ~0.357    & ~0.242    \\
\hline
$\frac{<A_L^2>-<A_L>^2}{<A_L^2>}$  &  0.1$\%$    &${\mathbf 0.3\%}$& $2\%$      &  $5\%$     & $13\%$   \\
\hline 
 \hline 
  \end{tabular}
\caption{\label{tab_mol1} The characteristic unpolarised cross sections and average asymmetries 
 for M{\o}ller  scattering calculated for $E_B = 250$ GeV.
 The boldface values are for the optimal working range marked in Fig. \ref{p_correl}.}
\end{center}
\end{table}
\begin{table}[ht]
\begin{center}
  \begin{tabular}[h]{|c|c|c|c|c|c|}
\hline
${\mathbf P_L^B\, \cdot \,P_L^T  }$    &${\mathbf \sigma}\ [\mu {\mathrm barn}]$ & ${\mathbf < A_R>}$
& ${\mathbf \Delta P_L^B}$/$ {\mathbf P_L^B}$   
    & ${\mathbf t_{Int}}$ {\bf [sec] }  & ${\mathbf N_{\mathrm \bf events}}$   \\
\hline\hline
0.8\,$\cdot$ \,0.5 & 0.274  & 0.292 &  1.0$\%$    & 2   &   $4 \cdot 10^5$    \\ 
\hline  
 0.8\,$\cdot$ \,0.2  & 0.342 & 0.117 &  1.0$\%$ & 12  &   $3 \cdot 10^6$    \\
\hline 
0.8\,$\cdot$ \,0.1 &  0.364& 0.058 &  1.0$\%$  & 53 &   $ 12\cdot 10^6$       \\
\hline
\hline
{ 0.8\,$\cdot$ \,0.08}   &{ 0.369}   &{ 0.047} & {  1.0}${ \%}$ & { 79}  &  
${ 18 \cdot 10^6}$   \\
\hline
 \hline 
  \end{tabular}
\caption{\label{tab_mol2} The characteristics 
and performance of a M\o ller polarimeter 
operated in the integrated polarisation measurement mode
calculated for $E_B = 250$ GeV , ~$\Delta P_L^T /P_L^T = 0.5 \%$,
an $x$ acceptance  in the region of $x_{min}\, - \,x_{max} = 0.66\, - \,1.34$
and $\Delta \phi \,= \, 40^o$. The values given are for $100 \% $ efficiency.}
\end{center}
\end{table}

\begin{table}[h]
\begin{center}
  \begin{tabular}[h]{|c||c|c|c||c|}
\hline
 &{\bf JLAB \cite{mol-jlab}} &{\bf E143 at SLC \cite{mol-e143}}&{\bf
 SLD at 
SLC \cite{swartz2}} & {\bf TESLA}   \\ 
\hline   \hline
Target & Fe & Fe+Co Alloy & Fe+Co Alloy & Fe \\
\hline
 $E_{B}$ [GeV]         & ~1 -- 6    & 16 $\ \&\ $ 29 & 46.6  &  250  \\
\hline 
 $\theta_{lab}$ [mrad] & ~32 -- 13  & 8 $\ \&\ $ 6   &  4.5   & 2.0 \\
\hline
$p_{lab}$ [GeV]        & 0.5 -- 3  &  ~~~8 $\ \&\ $ 14.5   &   23.3   & 125.0    \\
\hline
$(\Delta P_e)_{syst}$  & 0.5 $\%$ & 2.6 $\%$ & 3.4 $\%$ &
 0.85 $\%_{assumed}$     \\ 
\hline\hline
$\Delta P_e/P_e$ & $\simeq 1.3~\%$ & $\simeq 3.7~\%$ & $\simeq 4.2~\%$ & 1.0$\%$ \\
\hline 
  \end{tabular}
\caption{\label{mol-param2} 
A compilation of several relevant parameters of three existing 
M\o ller polarimeters, at  
high energy electron experiments, compared to a feasible
polarimeter configuration for TESLA. 
The $\theta_{lab}$ and $p_{lab}$ values are
calculated for $\theta=90^o$. The precision of the 
longitudinal beam polarisation
measurements $\Delta P_e/P_e$ is also listed. The value
of $\Delta P_e/P_e$ = 1.0 $\%$ given to 
TESLA corresponds to a measurement duration of 80 sec (see
Table \ref{tab_mol2})
assuming a 0.85 $\%$ systematic error.}
\end{center}
\end{table}


\end{document}